\documentclass[11pt,tightenlines,superscriptaddress,nofootinbib]{revtex4PATCH}

\usepackage[page,title]{appendix}
\usepackage{url}
\usepackage{placeins}
\usepackage{amsfonts}
\usepackage[dvipsnames]{xcolor}
\usepackage{color}
\usepackage{xspace}
\usepackage{tikz}
\usepackage[utf8]{inputenc}											
\usepackage{subfigure}

\usetikzlibrary{arrows}
\usetikzlibrary{shapes}
\usetikzlibrary{trees}
\usetikzlibrary{matrix}
\usetikzlibrary{arrows} 			
\renewcommand{\draft}[1]{{\color{blue}#1}}

\usepackage{graphicx,psfrag,bm}
\usepackage{mathrsfs,amsfonts}
\usepackage{epsfig,cancel}
\usepackage{epstopdf}
\usepackage{mciteplus}
\usepackage{latexsym}
\usepackage{amsthm}
\usepackage{amsmath}
\usepackage{amssymb}
\usepackage{hepunits}
\usepackage{hyperref}
\usepackage{bbm}
\usepackage{xfrac}
\usepackage{colordvi}
\usepackage{comment}
\usepackage{dcolumn}
\usepackage{times,wrapfig}
\usepackage{slashed}
\usepackage{booktabs}
\usepackage{relsize}

\newcommand{\be}{\begin{eqnarray}}
\newcommand{\ee}{\end{eqnarray}}

\newcommand{\benum}{\begin{enumerate}}
\newcommand{\eenum}{\end{enumerate}}
\newcommand{\bi}{\begin{itemize}}
\newcommand{\ei}{\end{itemize}}

\def\babar{\mbox{\slshape B\kern-0.1em{\smaller A}\kern-0.1emB\kern-0.1em{\smaller A\kern-0.2em R}}}

\newcommand{\brac}[2]{ \left( \frac{#1}{#2} \right) }

\newcommand{\Eq}[1]{Eq.~(\ref{#1})}

\colorlet{RED}{red}
\colorlet{BLUE}{blue}
\colorlet{ORANGE}{orange}

\begin{document}

\title{New Searches for Muonphilic Particles at Proton Beam Dump Spectrometers}

\date{\today}
\hspace*{0pt}\hfill 
{\small  FERMILAB-PUB-22-625-PPD-T}

\bigskip
\author{Diana~Forbes}
\affiliation{Department of Physics, University of Illinois Urbana-Champaign, Champaign, IL 61801, USA}

\author{Christian Herwig}
\affiliation{Fermi National Accelerator Laboratory, Batavia, IL 60510, USA}

\author{Yonatan~Kahn}
\affiliation{Department of Physics, University of Illinois Urbana-Champaign, Champaign, IL 61801, USA}
\affiliation{Illinois Center for Advanced Studies of the Universe, University of Illinois Urbana-Champaign, Urbana, IL 61801, USA}

\author{Gordan~Krnjaic}
\affiliation{Fermi National Accelerator Laboratory, Batavia, IL 60510, USA}
\affiliation{University of Chicago, Kavli Institute for Cosmological Physics, Chicago IL, USA}
\affiliation{University of Chicago, Department of Astronomy and Astrophysics, Chicago IL, USA}

\author{Cristina Mantilla Suarez}
\affiliation{Fermi National Accelerator Laboratory, Batavia, IL 60510, USA}

\author{Nhan~Tran}
\affiliation{Fermi National Accelerator Laboratory, Batavia, IL 60510, USA}

\author{Andrew~Whitbeck}
\affiliation{Texas Tech University, Lubbock, TX 79409, USA}

\begin{abstract}
We introduce a new search strategy for visibly decaying muonphilic particles using a proton beam spectrometer modeled after the SpinQuest experiment at Fermilab. In this setup,
a ${\sim}$100 GeV primary proton beam impinges on a thick fixed target and yields a secondary muon beam. As these muons traverse the target material, they scatter off nuclei and can radiatively produce
 hypothetical muonphilic particles as initial- and final-state radiation. If such new states decay to dimuons, their combined invariant mass can be measured with a downstream spectrometer immersed in a Tesla-scale magnetic field. For a representative 
 setup with $3\times 10^{14}$ muons on target with typical energies of $\sim$ 20 GeV, a $15\%$ invariant mass resolution, and an effective 100 cm target length,
 this strategy can probe the entire parameter space for which $\sim$ 200 MeV -- GeV scalar particles resolve the muon $g-2$ anomaly. 
 We present sensitivity to these scalar particles at the SpinQuest experiment where no additional hardware is needed and the search could be parasitically executed within the primary nuclear physics program.  
 Future proton beam dump experiments with optimized beam and detector configurations could have even greater sensitivity.  
\end{abstract}

\maketitle
\tableofcontents

\section{Introduction}

In recent years, searches for muonphilic particles -- new particles beyond the Standard Model that couple primarily to muons -- have attracted considerable interest, inspiring novel strategies involving
beam dumps \cite{Chen:2017awl}, $B$-factories \cite{BaBar:2016sci},  missing energy and momentum experiments \cite{Sieber:2021fue,Kahn:2018cqs}, the Large Hadron Collider
\cite{Galon:2019owl}, and even future muon colliders \cite{Capdevilla:2020qel} (see Ref. \cite{Capdevilla:2021kcf} and 
references therein for a survey of such techniques). In part, this popularity is related to the 
possible evidence for new physics from the Fermilab Muon $g-2$ collaboration, which has recently measured the anomalous magnetic moment of the muon ~\cite{Muong-2:2021vma,Muong-2:2021ojo,Muong-2:2021ovs,Muong-2:2021xzz}. This new result is consistent with the earlier Brookhaven measurement ~\cite{Muong-2:2006rrc} and the world average for $a_\mu \equiv \frac{1}{2} (g-2)_\mu$ now deviates from the Standard Model (SM) prediction ~\cite{Aoyama:2020ynm,Aoyama:2012wk,Aoyama:2019ryr,Czarnecki:2002nt,Gnendiger:2013pva,Davier:2017zfy,Keshavarzi:2018mgv,Colangelo:2018mtw,Hoferichter:2019mqg,Davier:2019can,Keshavarzi:2019abf,Kurz:2014wya,Melnikov:2003xd,Masjuan:2017tvw,Colangelo:2017fiz,Hoferichter:2018kwz,Bijnens:2019ghy,Colangelo:2019uex,Pauk:2014rta,Danilkin:2016hnh,Jegerlehner:2017gek,Knecht:2018sci,Eichmann:2019bqf,Roig:2019reh,Colangelo:2014qya} by 
\be \label{amu-exp}
\Delta a_\mu  =  a_{\mu}^{\rm exp}- a_{\mu}^{\rm theory} = (251 \pm 59) \times 10^{-11}~,
\ee
which constitutes a statistically significant $4.2 \sigma$ discrepancy.\footnote{Previous lattice QCD extractions of the hadronic light-by-light \cite{Gerardin:2019vio,Blum:2019ugy,Chao:2021tvp} and hadronic vacuum polarization \cite{FermilabLattice:2017wgj,Budapest-Marseille-Wuppertal:2017okr,RBC:2018dos,Giusti:2019xct,Shintani:2019wai,FermilabLattice:2019ugu,Gerardin:2019rua,Aubin:2019usy,Giusti:2019hkz} contributions to $a_\mu$ are consistent with both the measured value and semi-analytical calculations based on $R$-ratio data. However, the BMW collaboration \cite{Borsanyi:2020mff} has extracted a SM prediction of $a_\mu$ that is consistent with the measured value. This result is in tension with $a_\mu$ as determined by $R$-ratio methods and might be in tension with the SM electroweak fit \cite{Passera:2008jk,Crivellin:2020zul,Keshavarzi:2020bfy}, so future lattice calculations and improved $R$-ratio data will be necessary to conclude whether \Eq{amu-exp} is evidence of new physics.}
 
If this discrepancy is due to new physics, there are necessarily new particles in nature that couple to muons. These particles can be classified according to whether they are heavy with order-unity couplings to the muon (e.g. new weak-scale states charged under SM gauge interactions) or light and feebly coupled. If the former possibility is realized in nature, such states could be discovered with future high-energy collider searches~\cite{Capdevilla:2021rwo,Capdevilla:2020qel}. If instead, the particles in question are light and feebly coupled, the possibilities for new physics are much narrower: the particles must be muonphilic scalars or vectors, which are singlets under the SM gauge group \cite{Capdevilla:2021kcf}. It has been shown that, for most available decay channels, existing and planned intensity frontier experiments have sufficient sensitivity to discover these light new states if their couplings to the muon resolve $\Delta a_\mu$~\cite{Kahn:2018cqs,Chen:2017awl,Berlin:2018pwi,Capdevilla:2021kcf,Sieber:2021fue,Coy:2021wfs,Rella:2022len}. However, there is a notable exception to this otherwise comprehensive coverage: scalars $S$ with sub-GeV masses and prompt dimuon decays.

In this paper, we propose muon spectrometers at proton beam dumps as promising experiments to search for $S \to \mu^+ \mu^-$ and potentially discover the new physics responsible for $\Delta a_\mu$. While the parameter space of interest in this paper is framed around scalar particles that resolve $g-2$, the search strategy we present is general
and can be adapted to search for any new particles that decay appreciably to dimuons. Proton beam dumps feature enormous luminosity and copious secondary production of muons through pion decay, evading some of the event rate limitations that cap the sensitivity of muon beam experiments such as M$^3$ \cite{Kahn:2018cqs} and NA64-$\mu$ \cite{Sieber:2021fue}. The production of $S$ from bremsstrahlung during nuclear scattering in the dump, $\mu^\pm N \to \mu^\pm N S$, is only weakly dependent on the muon energy, and thus a mono-energetic muon beam is not necessary. As we will show, for $S$ couplings that resolve $\Delta a_\mu$, the $S$ decay is prompt and the signal is an invariant mass peak in opposite-sign muons emerging from a single vertex in the dump. The sensitivity is therefore driven by the invariant mass resolution, and we will argue that selecting events in the final portion of the dump, combined with a high-momentum-resolution spectrometer magnet, suffices to achieve the invariant mass resolution needed to observe a signal above the SM background from continuum QED production. We emphasize that our proposed search strategy involves no new hardware and can be parasitically implemented at any proton beam spectrometer experiment. 

This paper is organized as follows.  We introduce the scalar singlet model in Section~\ref{sec:theory} as a solution to $\Delta a_\mu$, which can also be seen as a representative example of a muonphilic model. In Section~\ref{sec:concept}, we introduce the basic experimental concept and discuss the generic requirements on the detector configuration.  Then, in Section~\ref{sec:calculations}, we calculate the scalar singlet production rate as well as that of the leading irreducible background processes, and propose cuts which maximally exploit the different kinematics of signal and background. In Sec.~\ref{sec:experiment}, we take the specific case of the SpinQuest experiment at Fermilab and detail key characteristics of the experiment, expected backgrounds, and the potential sensitivity to muonphilic scalars.  Finally, we conclude and provide an outlook on the near-term experimental prospects in Section.~\ref{sec:outlook}.


\section{Scalar Singlet Model}
\label{sec:theory}

We extend the SM by a scalar $S$ which is a singlet under the SM gauge group, and which couples exclusively to muons through the Yukawa interaction 
\be
\label{eq:LS}
\mathcal{L} \supset g_S S \, \overline{\mu} \mu~,
\ee
where $g_S$ is a dimensionless coupling constant. 
This coupling induces a shift in $\Delta a_\mu$ at one-loop level which yields
\begin{equation}
\label{eq:aS}
\Delta a^{S}_{\mu}  =    \frac{g_S^2}{8\pi^2} \int_0^1 dz \frac{ (1+z) (1-z)^2}{(1-z)^2 +  z (m_S/m_\mu)^2 } \approx 2 \times 10^{-9}  \, \left(\frac{g_S}{10^{-3}}\right)^2   \left( \frac{700 \, \rm MeV}{  m_S} \right)^{2}~~,
\end{equation}
where $m_S$ is the mass of $S$. The approximate equality holds in the limit $m_S \gg m_\mu$, and gives a sense for the typical size of couplings required. In the absence of other interactions, for $m_S > 2 m_\mu$ the only tree-level decay channel is $S \to \mu^+ \mu^-$, with a corresponding width of
\begin{equation}
\label{eq:width}
    \Gamma(S \rightarrow \mu^+ \mu^-) = \frac{g_S^2 m_S}{8 \pi}\left(1-\frac{4 m_\mu^2}{m_S^2} \right)^{3/2}~.
\end{equation}
At one-loop level, there are also $S \to \gamma \gamma$ and $S \to \bar \nu \nu$ decay channels, but these are further suppressed by $\alpha^2$ and $G^2_F$, respectively, so the dimuon channel has a branching fraction near unity. 

As noted in Ref.~\cite{Batell:2017kty,Batell:2021xsi}, the interaction in Eq.~(\ref{eq:LS}) is not gauge-invariant under the electroweak symmetry of the SM, but it may be generated from the dimension-5 operator 
\be
{\cal L}_{\rm UV} = \frac{1}{\Lambda} S H^\dagger L_\mu \mu^c  + {\rm h.c.}~,
\ee
where, in two-component Weyl fermion notation, $L_\mu$ is the second generation lepton doublet, $\mu^c$ is the right handed muon field, $H$ is the SM Higgs doublet, and $\Lambda$ is the mass scale at which new particles have been integrated out. Matching Wilson coefficients yields $g_S  = v/\Lambda$, so for $g_S = 10^{-3}$, the cutoff scale of the effective field theory defined by~\Eq{eq:LS} is valid up to $\Lambda \approx 250 \ {\rm TeV}$. Consequently, for the parameter space we consider in this paper ($m_S \lesssim 5 \ {\rm GeV}$ and muon beam energies below 50 GeV), the effective theory description is perfectly valid. 

Previous work to constrain this model has focused on search strategies to probe the invisible decay channel if $S$ couples to neutrinos or dark matter \cite{gninenko:2016kpg,Kahn:2018cqs,Krnjaic:2019rsv}, as well as other loop-suppressed modes for $m_S < 2m_\mu$ which lead to long-lived $S$ and displaced diphoton decays through one-loop processes~\cite{Chen:2017awl}. By contrast, here we focus on the visible decay mode $S \to \mu^+ \mu^-$, which we assume has a 100\% branching fraction for $m_S > 2m_\mu$. Allowing for alternate invisible decay modes (which may imply the presence of new dark states) reproduces the phenomenology of previous invisible decay studies such as M$^3$~\cite{Kahn:2018cqs} and NA64-$\mu$~\cite{gninenko:2016kpg} and we will not consider these in this paper.

Finally, we note that the phenomenological search strategy introduced below may also be applied to new spin-1 vector particles whose couplings to the muon resolve $\Delta a_\mu$. However, nearly all theoretically consistent models for visibly-decaying vectors in this mass range have been ruled out by laboratory searches~\cite{Bauer:2018onh,Capdevilla:2021kcf}. The only remaining anomaly-free $U(1)$ extension to the SM that can still resolve $\Delta a_\mu$ is gauged $L_\mu - L_\tau$ \cite{Bauer:2018onh} for vector masses between $\sim 10$-200 MeV, where the lower bound is set by cosmology 
\cite{Escudero:2019gzq} and the upper bound is set by the BABAR $e^+e^- \to 4 \mu$ search \cite{BaBar:2016sci}. Thus, for nearly all of the remaining viable parameter space in this model, the vector particle decays invisibly to neutrinos and is, therefore, testable with NA64-$\mu$ \cite{Gninenko:2019qiv} and M$^3$ \cite{Kahn:2018cqs}, which are optimized for missing momentum signatures. By contrast, scalar particles that resolve $\Delta a_\mu$ can still visibly decay to dimuons \cite{Capdevilla:2021kcf}, so we focus on this scenario throughout our analysis.

\section{Proton Beam Dump Spectrometer Concept}
\label{sec:concept}

\begin{figure}[t!]
    \centering
    \includegraphics[width=0.8\textwidth]{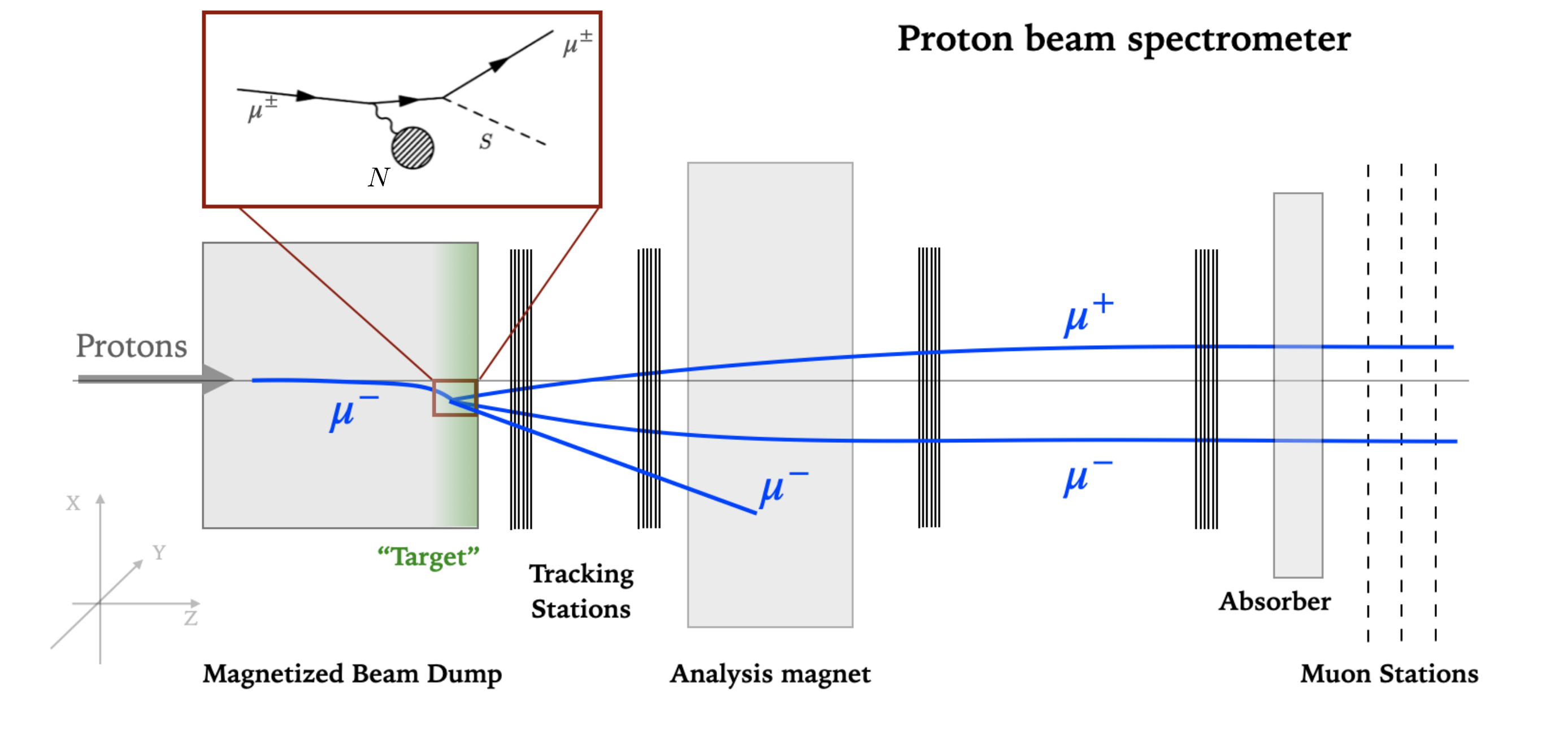}
    \caption{
     Proton beam dump spectrometer signature of prompt muonphilic scalars produced in the back of the beam dump, labeled as the ``Target'', and reconstructed by the downstream tracking stations. The spectrometer setup is inspired by the existing SpinQuest experiment, but we argue that the search strategy presented in this paper can work for other proton beam spectrometer configurations with a large flux of muons (see main text for details).}
    \label{fig:signature}
\end{figure}

To search for muonphilic scalars as a possible explanation for $\Delta a_\mu$, we require a large flux of muons on a target which will produce the scalars via bremsstrahlung, as shown in Fig.~\ref{fig:signature} (red boxed inset). 
From the decay width of the $S$ to muons in \Eq{eq:width}, the lab-frame decay length is:
\begin{equation}
    L \approx 8 \times 10^{-8} \, {\rm m} \brac{E_S/m_S}{10}  \left( \frac{700 \, \rm MeV}{  m_S} \right)\left(\frac{10^{-3}}{g_S}\right)^2~,
\end{equation}
where we have taken the $m_S \gg m_\mu$ limit. Except for a very small region of phase space just above the dimuon threshold, the couplings required to explain $\Delta a_\mu$ imply that the $S$ must decay promptly. This remains true even if there are additional invisible decay modes, since those will only increase the total width and hence decrease the decay length.

Therefore, the target itself cannot be very dense or else the momentum resolution will be degraded by multiple scattering, so a large muon flux is important to compensate for this lower density.
In this paper, we consider the SpinQuest spectrometer as an example of an experimental setup to search for such muonphilic scalars. A schematic inspired by SpinQuest is shown in Figure~\ref{fig:signature}. The proton beam travels through some magnetized material producing a large fraction of $\mu^\pm$ with $\mathcal{O}(20 \ {\rm GeV})$ energies, most of which originate from pion decays. 
These secondary muons are produced along the beam dump and those muons traversing the target region, which in Fig.~\ref{fig:signature} is denoted by ``Target'' in green, can produce the $S$ during a nuclear scattering event and the outgoing daughter muons have enough momentum to exit the dump and be detected.  
The path of the beam muon is deflected by the magnetized dump, while the analysis magnet alters the trajectories of the three outgoing muons to measure their curvature and hence momenta. The signal is thus two or three muons originating from the same vertex in the dump (depending on whether the third muon has a high enough momentum to emerge from the tracking stations), with the invariant mass of an opposite-sign muon pair reconstructing the mass of the $S$. We note that Ref.~\cite{Berlin:2018pwi} previously considered $S$ production from secondary muons at SpinQuest but focused on the $e^+ e^-$ decay mode, whereas we focus on the irreducible decay to $\mu^+ \mu^-$.

This search strategy can be employed at any proton beam dump spectrometer with the following key features: 
\begin{itemize}
    \item a high-intensity proton beam with high repetition rate, from which a large flux of muons is produced.  A high repetition rate, near-continuous wave beam is more valuable than a beam of similar current but lower duty factor (pulsed) in order to reduce combinatorial backgrounds;
    \item a beam dump that is many nuclear interaction lengths thick, to allow predominantly muons to exit through through the dump and greatly reduce hadronic backgrounds, with the last portion of the dump serving as the target;
    \item a beam of sufficient energy to produce secondary muons capable of traversing the entire dump, retaining sufficient momenta to both produce the $S$ and to boost its decay products into the spectrometer's acceptance;
    \item a magnetized beam dump to spatially spread out positive and negatively charged beam muons to reduce the combinatorial backgrounds;
    \item a detector that can trigger on dimuon coincidences in the presence of a high muon flux, has good momentum and angular resolution such that the invariant mass resolution is mostly influenced by multiple scattering, and is granular enough to reject backgrounds.
\end{itemize}

As we will see below, the dominant irreducible background is muon pair production from QED, while an important reducible background is the combinatorial background from independent production of muons in the dump. Because the search strategy is a bump hunt in invariant mass, the relative size of the QED background will be driven both by the invariant mass resolution and the effective target length $\ell_T$ (the green region in Fig.~\ref{fig:signature}). These requirements are in some tension because a larger $\ell_T$ improves the sensitivity (which scales as $\sqrt{\ell_T}$ assuming Poisson fluctuations on the background), but also leads to increased multiple scattering and a degraded mass resolution, as well as a larger combinatorial background. A full optimization of the sensitivity with respect to $\ell_T$ requires a concrete experimental design and is beyond the scope of this study, but as an example, we will take $\ell_T = 100 \ {\rm cm}$ and a 15\% invariant mass resolution. This figure represents the combined effects of an intrinsic 5\% experimental resolution and multiple scattering and is further justified with simulations in Sec.~\ref{sec:massres}.

\section{Signal and Background Rates, Kinematics, and Cuts}
\label{sec:calculations}

\subsection{Signal and irreducible background}

\begin{figure}[t!]
\centering
\includegraphics[width=0.7\textwidth]{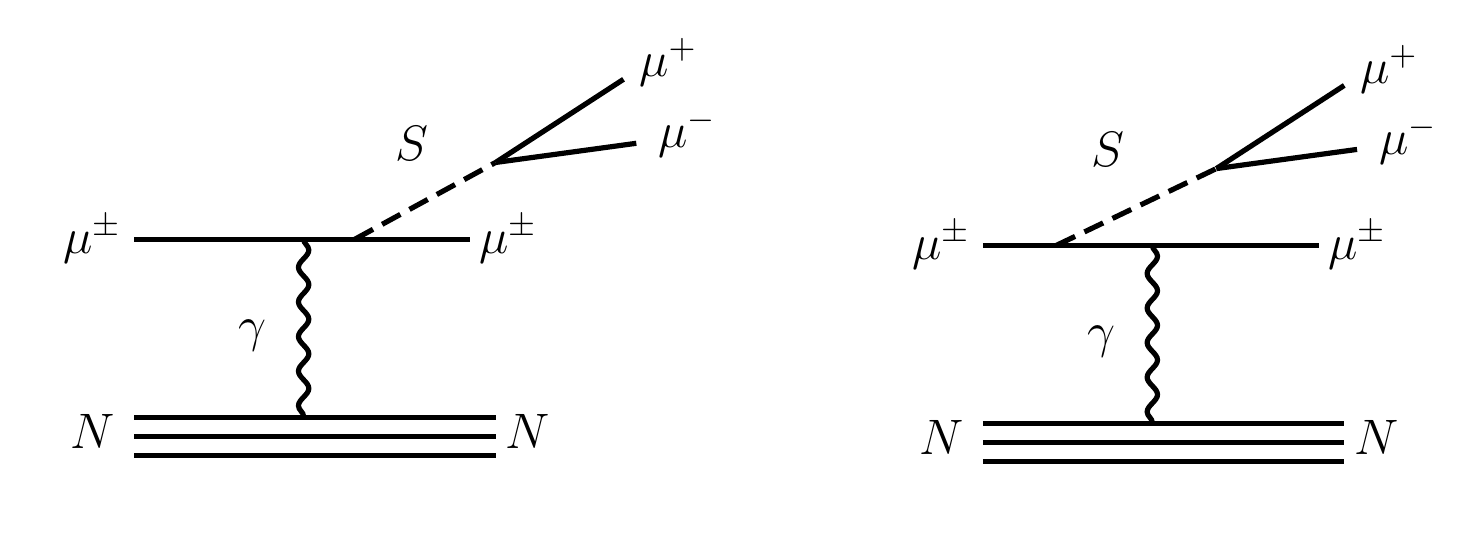}
\caption{Feynman diagrams representing
the dominant signal processes for $S$ production
in muon-nucleus scattering.
\label{fig:Signal}
}
\end{figure}

\begin{figure}[t]
\centering
\includegraphics[width=0.7\textwidth]{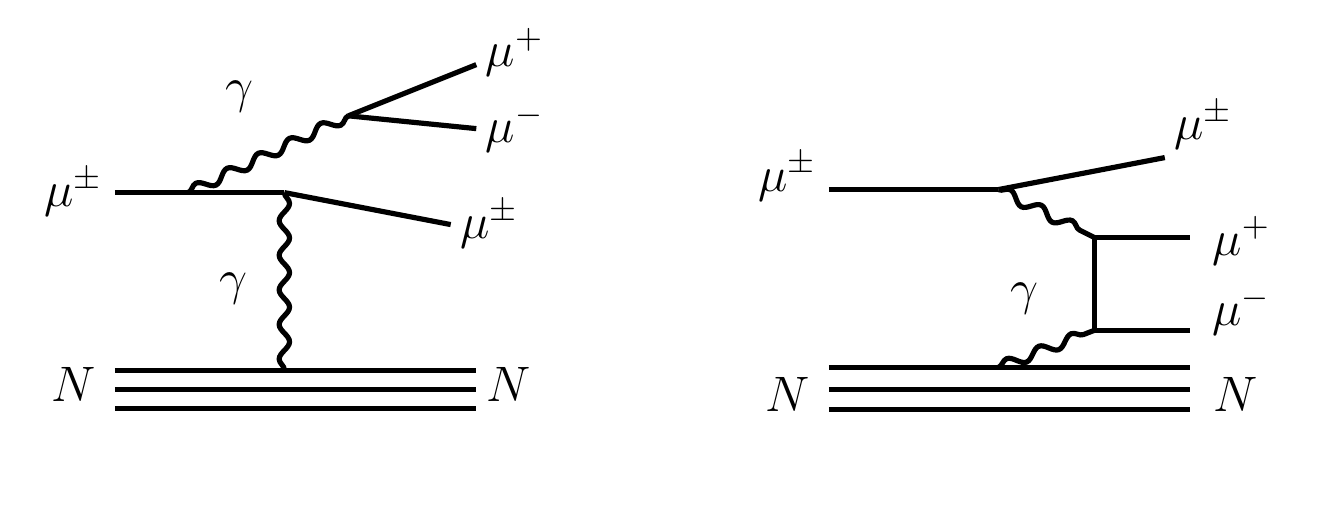}
\caption{Feynman diagrams representing
the dominant Standard Model background processes for dimuon pair production
in muon-nucleus scattering. The radiative diagram (left) also has a contribution from final-state radiation, but radiation off the nucleus is suppressed since $m_N \gg m_\mu$.
\label{fig:SMBackground}
}
\end{figure}

Our signal process is on-shell $S$ production from muons scattering off a fixed target of nuclei $N$, $\mu^\pm N \to \mu^\pm N S$, followed by the prompt $S \to \mu^+ \mu^-$ decay (Fig.~\ref{fig:Signal}). The experimental signature includes at least two opposite-sign muons in the final state originating from a single vertex which reconstruct the invariant mass of the $S$. The SM background is dominantly continuum muon pair production $\mu^\pm N \to \mu^\pm N \mu^+ \mu^-$ through electromagnetic processes, including both the radiative (Fig.~\ref{fig:SMBackground}, left) and Bethe-Heitler trident (Fig.~\ref{fig:SMBackground}, right) diagrams \cite{Ganapathi:1980wh,Bjorken:2009mm}. The background from off-shell $Z$ production is negligible at the beam energies we consider. Photon-initiated background processes resulting from a hard photon bremsstrahlung, $\mu N \rightarrow \mu N \gamma$ followed by  $\gamma N \rightarrow \text{hadrons}$, are negligible since they can be mitigated with a mild di-muon selection and are further suppressed by $(m_e/m_\mu)^2$ for muon beams compared to the case of electron beams.

We simulated both signal and background processes with CalcHEP~\cite{Belyaev:2012qa}. Although MadGraph~\cite{Alwall:2011uj} has become a standard tool for fixed-target experiments as well as collider experiments, it has known issues with collinear emission processes
\cite{Chen:2017awl} which give large fluctuations in the computed cross section, while CalcHEP reliably gives sub-percent level Monte Carlo errors~\cite{Marsicano:2018vin}. We implemented a custom form factor in CalcHEP for the nuclear target by scaling all the cross sections by a dipole form factor~\cite{Essig:2010xa}:
\be
    G_2(t) = G_2^{\rm el}(t) + G_2^{\rm inel}(t)~~,
\ee
where the elastic contribution is 
\be
    G_2^{\rm el}(t) = Z^2 \left(\frac{a^2 t}{1 + a^2 t} \right)^2 \left(\frac{1}{1 + t/d} \right)^2,
\ee
and the inelastic contribution can be written as
\be
G_2^{\rm inel}(t) = Z \left(\frac{a'^2 t}{1 + a'^2 t} \right)^2  W_2(t)~~,~~~   W_2(t) = \left[\frac{1 + \tau (\mu_p^2 - 1)}{\left(1+t/t_0\right)^4} \right]^2~~.
\ee
The parameters of this form factor model are
\be
    a = \frac{113 Z^{-1/3}}{m_e}, ~~~~~ a' = \frac{773 Z^{-2/3}}{m_e},  ~~~~~ d = 0.164 \ \text{GeV}^2 A^{-2/3},~~~~ t_0 = 0.71 \text{  GeV}^2~~,
\ee
where $\mu_p = 2.79$, $\tau = t/(4 m_p^2)$, $Z$ and $A$ are the atomic number and mass number of $N$, $m_p$ and $m_e$ are the proton and electron masses, and $t = -(p_N' - p_N)^2$ is the squared 4-momentum transfer to the nucleus (initial 4-momentum $p_N$, final momentum $p_N'$). For the signal events, in order to avoid any issues arising from the narrow width of the $S$, we generated on-shell $S$ events $\mu^\pm N \to \mu^\pm N S$, decayed the $S$ isotropically in its rest frame, and boosted back to the lab frame. In order to confirm that the form factor was correctly implemented in CalcHEP, we reproduced the results of Fig.\ 8 of Ref.~\cite{Marsicano:2018vin} for the process $\mu N \rightarrow \mu N S$ with $N$ an aluminum nucleus. The number of signal events or background events $N_{S,B}$ is given by
\be
N_{S,B} = \sigma^{(\rm acc)}_{S,B} \, n_T \, \ell_T \times {\rm MOT},
\ee
where $\sigma^{(\rm acc)}$ is the accepted cross section given any cuts applied (described further below), $n_T$ and $\ell_T$ are the number density of nuclei in the target and the effective target length, respectively, and MOT is the number of muons on target.

\subsection{Kinematics and cuts}

\begin{figure}[t!]
\centering
\includegraphics[width=0.5\textwidth]{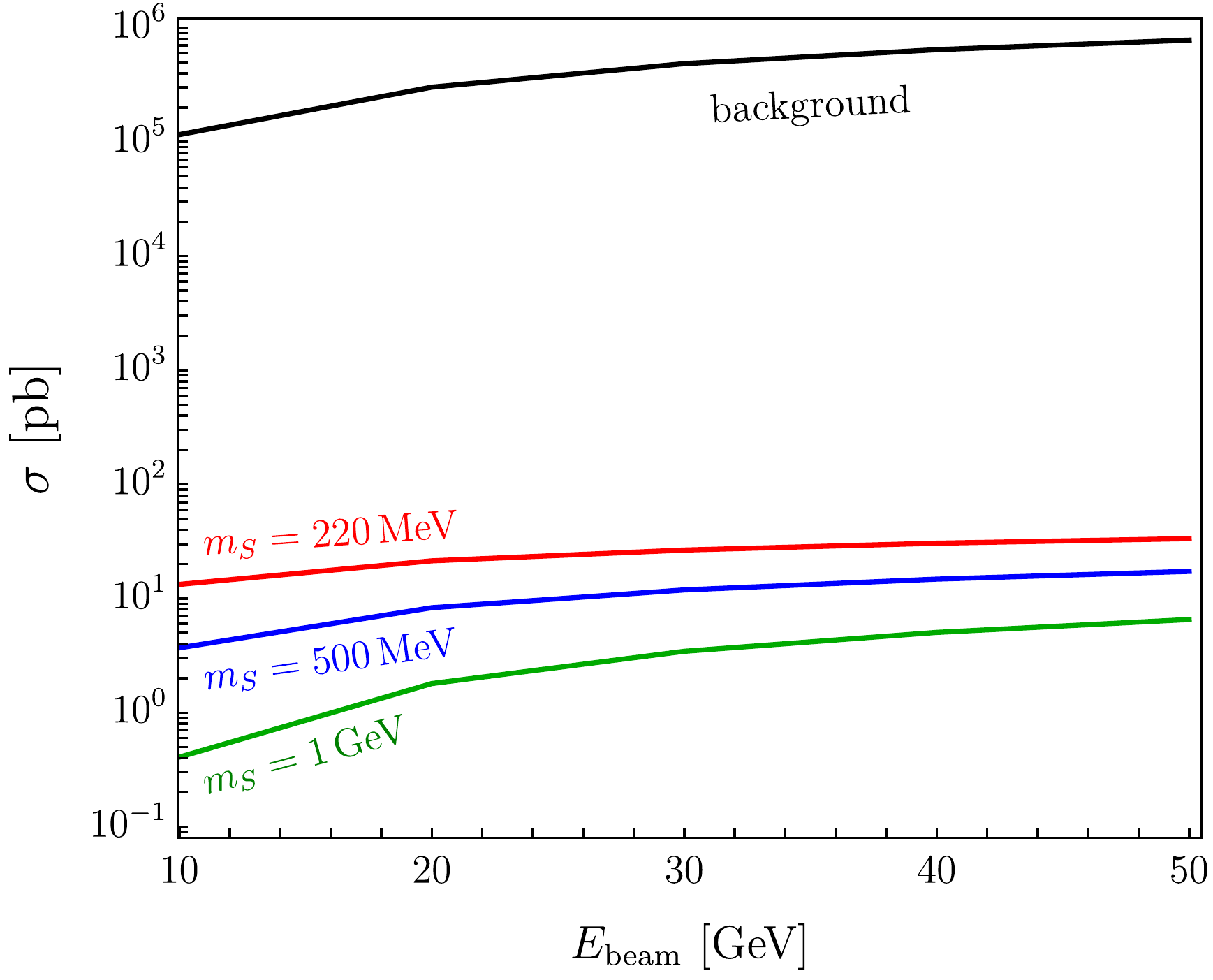}
\caption{Cross sections as a function of beam energy for background and signal for various $S$ masses, with $g_S$ set to the minimum value consistent with the measured $\Delta a_\mu$ to within $2\sigma$.}
\label{fig:xsec}
\end{figure}

As a starting point, based on the setup envisioned in Section~\ref{sec:experiment} below, we will consider the example of a $\mu^-$ beam on an iron target with $Z = 26$ and $A = 56$, $n_T = 8.5 \times 10^{22}/{\rm cm}^3$, and $\ell_T = 100 \ {\rm cm}$.
Figure~\ref{fig:xsec} shows, as a function of the beam energy, the cross section for $\mu^{-} \mu^{-}\mu^{+}$ events from the QED background process and signal process for different scalar masses, fixing the coupling $g_S$ for each mass to be the minimum coupling required to resolve $\Delta a_\mu$ to within $2\sigma$. The cross section is largely insensitive to the beam energy, but for all energies and masses, the QED background is several orders of magnitude larger, requiring additional cuts to render the signal visible. The most obvious such cut is an invariant mass cut; this turns out to be most efficient when performed on the opposite-sign muon pair with the \emph{hardest} $\mu^{-}$, which we can justify by comparing the signal and background kinematics.

\begin{figure}[t!]
\centering
\includegraphics[width=0.45\textwidth]{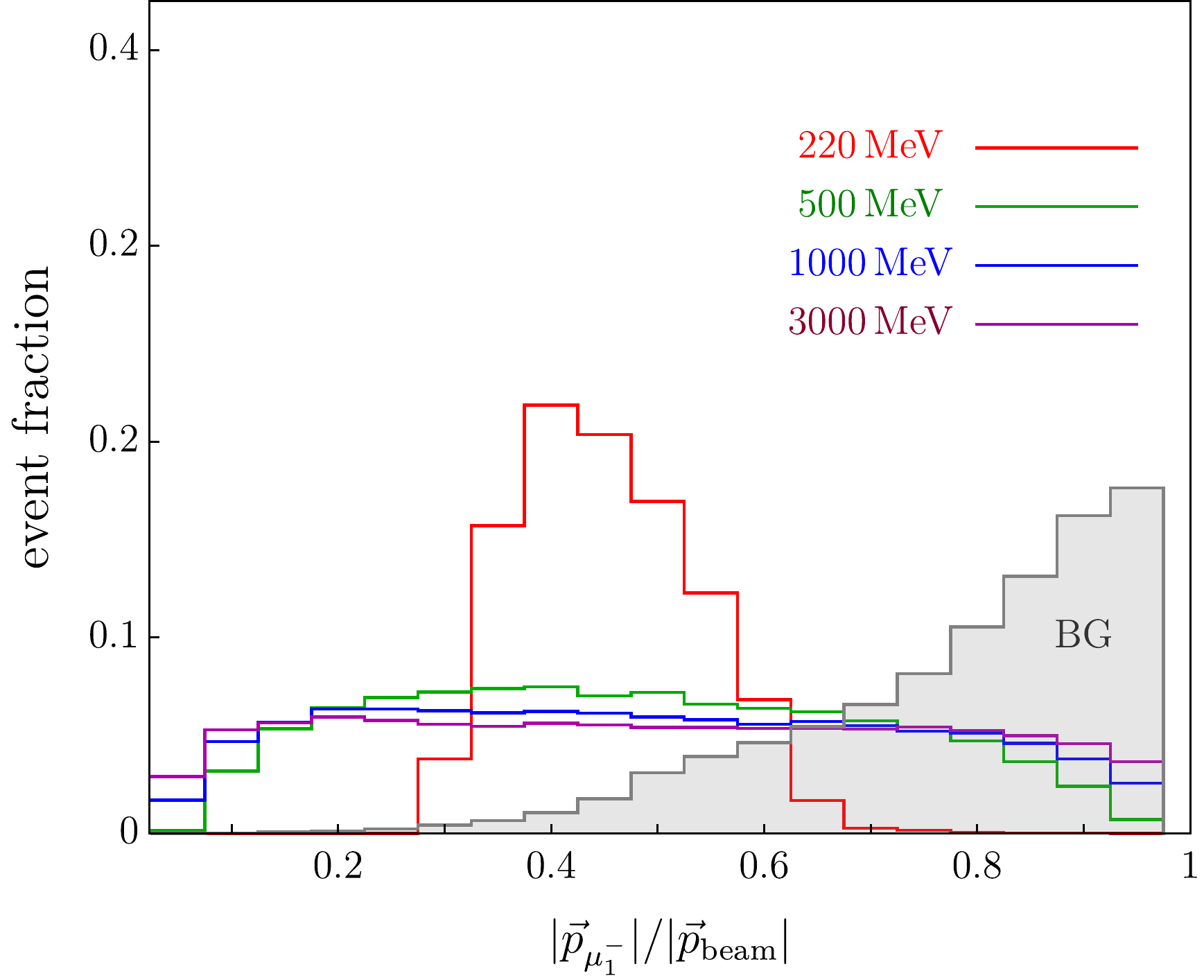}~~~
\includegraphics[width=0.45\textwidth]{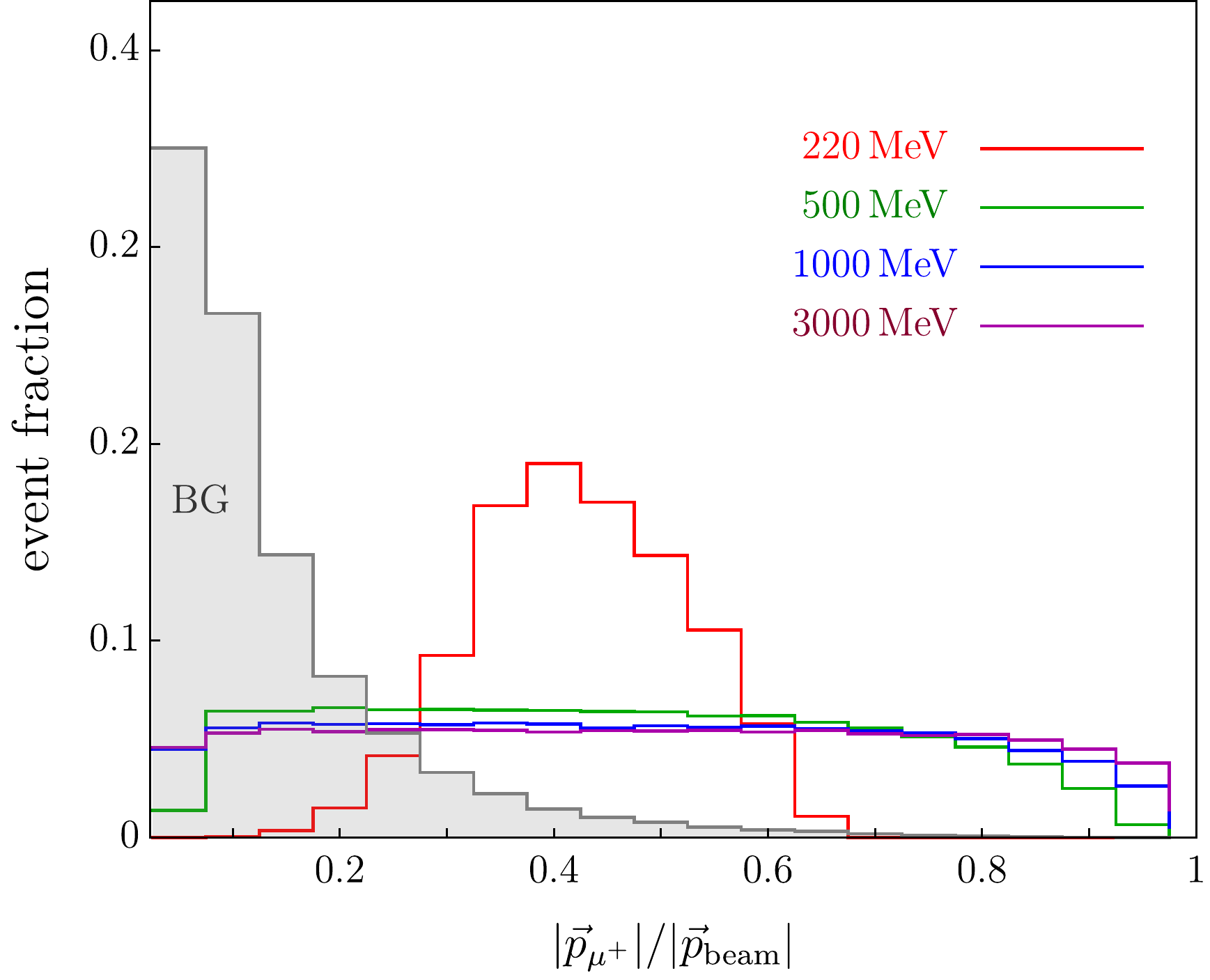}
\includegraphics[width=0.45\textwidth]{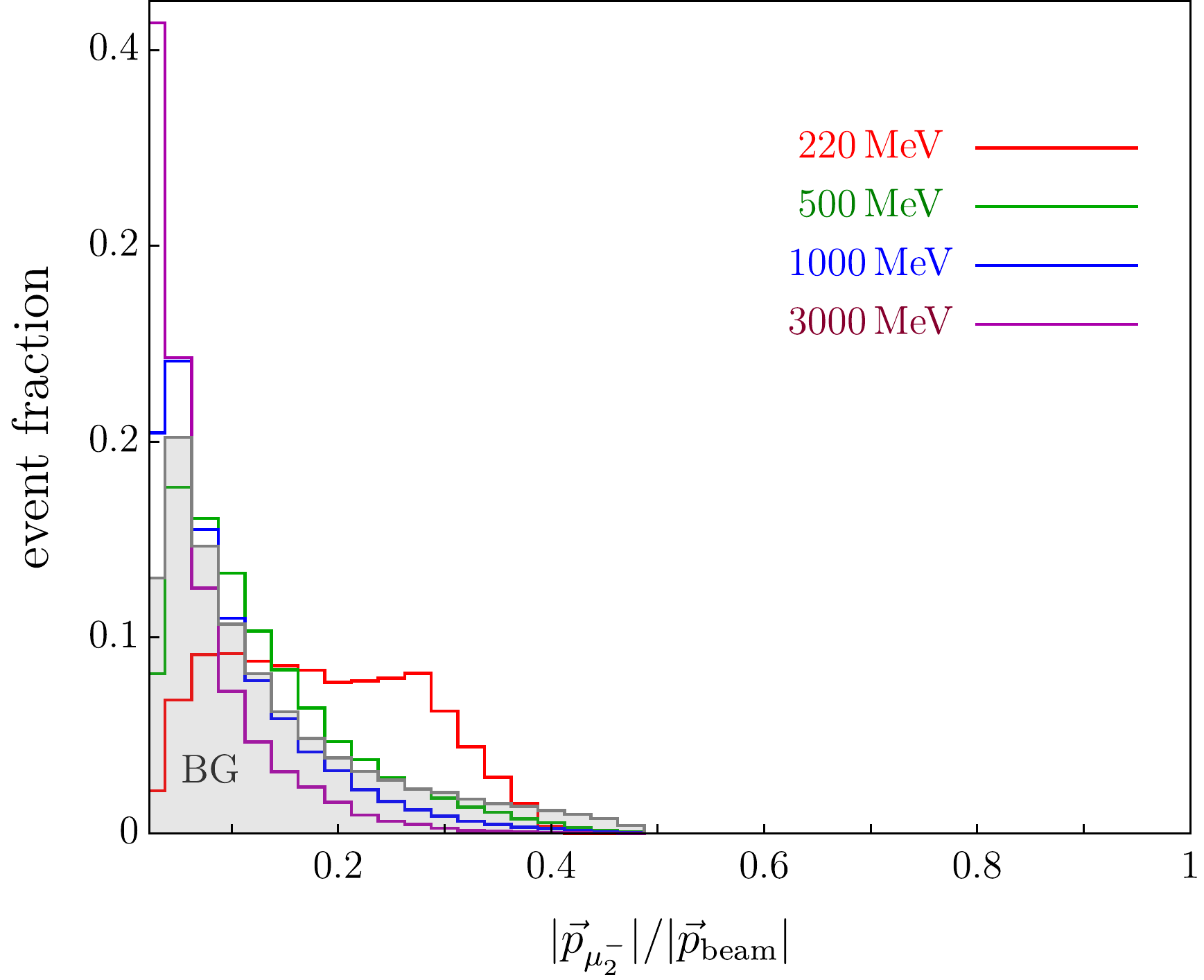}~~~
\includegraphics[width=0.46 \textwidth]{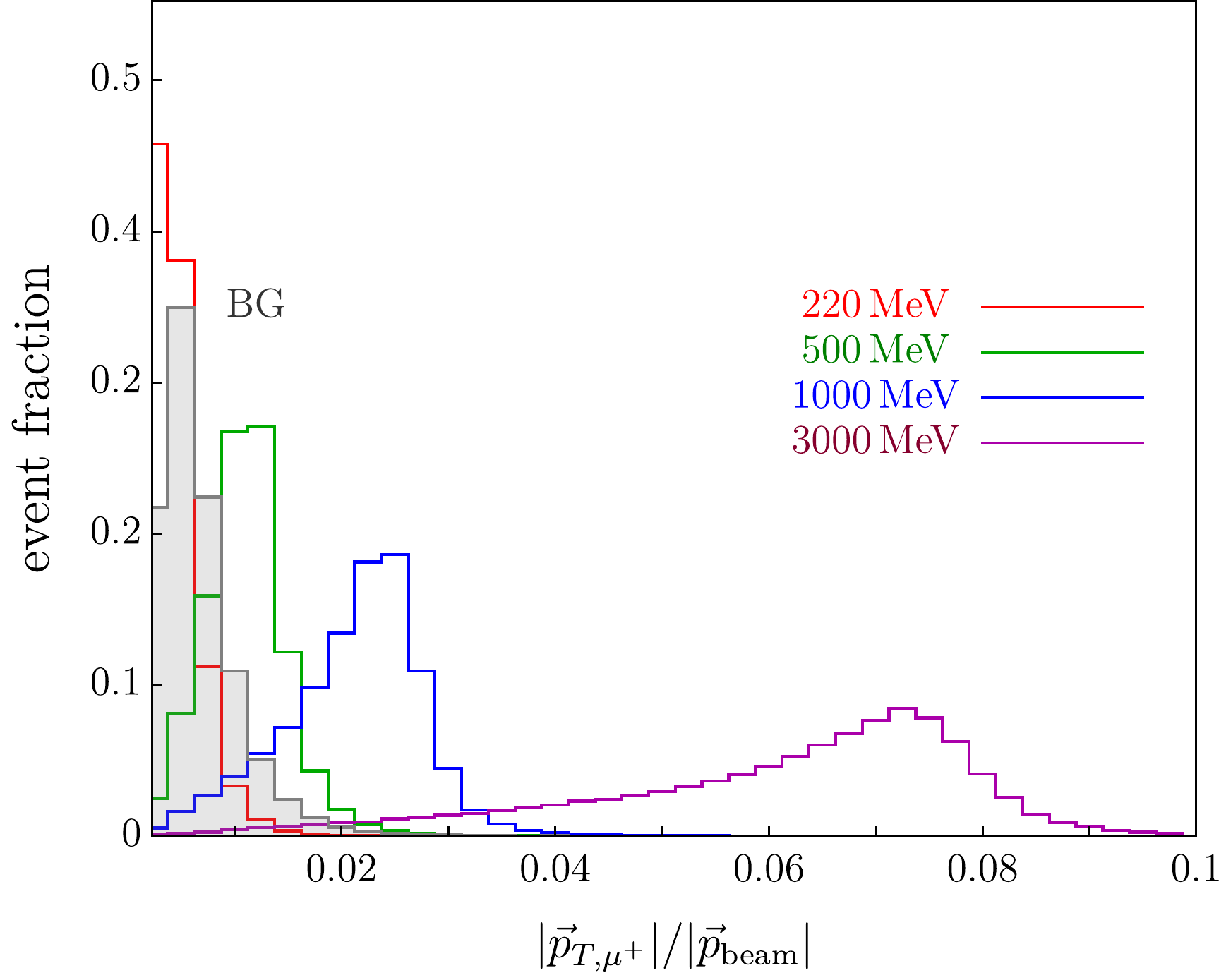}
\caption{Kinematic distributions for $\mu^{-}\mu^{-}\mu^{+}$ events produced by SM QED background (shaded gray, labeled ``BG'') and signals of various scalar masses with a 20\,GeV $\mu^{-}$ beam.
Relative distributions (normalized to unity for illustration) are shown for the fraction of the beam momentum carried by the hardest final-state $\mu^{-}$, labeled $\mu_1^{-}$ \textbf{(top left)}, $\mu^{+}$ \textbf{(top right)}, and softer $\mu^-$, labeled $\mu_2^{-}$ \textbf{(bottom left)}.
The momentum of the leading $\mu^{-}$ transverse to the beam is also shown \textbf{(bottom left)}. For heavy $S$, the $p_T$ and angular separation of the hardest muons peak at larger values than the background.
}
\label{fig:ch1}
\end{figure}

Figure~\ref{fig:ch1} shows the kinematic distributions of both signal and background events for a 20 GeV $\mu^{-}$ beam, without applying any cuts. The principal difference between the signal and background processes is that the cross section for $S$ production is peaked when $S$ takes nearly all the beam energy~\cite{Liu:2016mqv,Kahn:2018cqs}, at least for $m_S > m_\mu$ which is the case whenever $S$ can decay visibly. As a result, the $S$ is highly boosted, and thus the harder $\mu^-$ dominantly originates from $S$ decay, with the momenta of the $\mu^+$ and harder $\mu^-$ both peaked at $|\vec{p}_{\rm beam}|/2$. By contrast, the soft singularity of the massless photon in QED means that the $\mu^+$ momentum distribution is peaked at zero, while the harder $\mu^-$ is typically the beam muon which still carries a large fraction of the original beam momentum.\footnote{We note that the same distinctions between the QED and muonphilic particle kinematics hold for massive vectors, pseudoscalars, and axial vectors~\cite{Liu:2017htz}, and thus the same search strategies we propose should apply to those models too.} Both of these features are apparent in the top panel of Fig.~\ref{fig:ch1}. Combined with the fact that the high boost afforded by fixed-target experiments makes almost all events purely in the forward direction, a cut on $p_z$ for both the $\mu^+$ and hardest $\mu^-$ allows an efficient separation of signal and background. We also note that for both signal and background, the recoil $\mu^{-}$ is very soft and often carries less than 25\% of the beam momentum (bottom left). In practice, the soft $\mu^-$ may not even be observable if it curves away from the tracking stations, so absent any issues from combinatorial background due to the fact that the beam contains both $\mu^+$ and $\mu^-$, we can define the signal as two muons of opposite sign originating from the same vertex. We argue in Sec.~\ref{sec:experiment} below that the combinatorial background is likely negligible compared to the irreducible background for $m_S \lesssim 1 \ {\rm GeV}$. Finally, the bottom-right panel of Fig.~\ref{fig:ch1} shows the transverse momentum distributions of the $\mu^+$. The collinear singularity of QED (regulated by the muon mass) implies that the background peaks at small $p_T$, while for heavy $S$, the $p_T$ is peaked at larger values.\footnote{A cut on $p_T$ would only efficiently separate signal from background at masses $m_S \gtrsim 1 \ {\rm GeV}$, where (as we illustrate below) the irreducible QED background is already subdominant to the expected combinatorial background after the invariant mass and $p_z$ cuts.}

\begin{figure}[t!]
\centering
\includegraphics[width=0.48\textwidth]{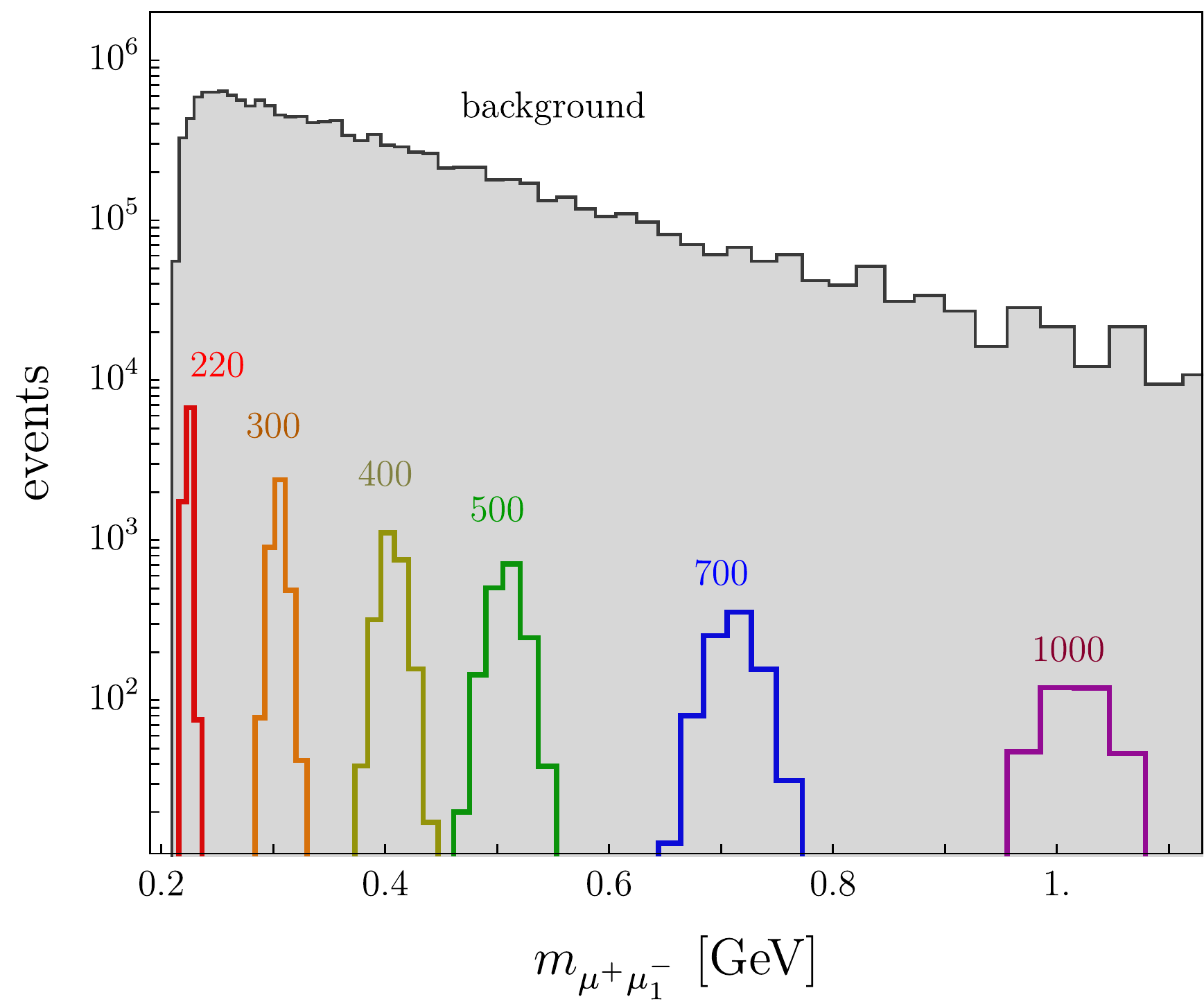}
\caption{The $\mu^{+}\mu^-$ invariant mass distribution is shown for background and signal for several scalar masses (in MeV units), in events produced from a 20\,GeV $\mu^{-}$ beam. 
A $5\times10^{13}$ MoT sample (corresponding to approximately 1 year of running at SpinQuest, see Sec.~\ref{sec:experiment}) is shown with $g_S$ set to the minimum values which resolve $\Delta a_\mu$ for each $m_S$ value, as in Fig.~\ref{fig:xsec}.
Here, the invariant mass is formed from the harder $\mu^-$ (labeled $\mu^-_1$) in the event and 3-momenta are smeared to produce a 15\% mass resolution. Each muon is required to satisfy $p_z>5$\,GeV.
\label{fig:masspre}
}
\end{figure}

Fig.~\ref{fig:masspre} shows the invariant mass spectrum of the $\mu^{+}$ paired with the hardest $\mu^{-}$, for both signal and background, after applying the $p_z$ cut. As in Fig.~\ref{fig:xsec}, we have normalized the signal to the values of $g_S$ consistent with $\Delta a_\mu$. Keeping in mind the possible degradation of the mass resolution with multiple scattering in the target, we take as an example a $15 \%$ invariant mass cut, which means we require that the harder muon pair has an invariant mass within $m_S \pm r$, where $2r/m_S = 0.15$. In Fig.~\ref{fig:masspre} we have applied a uniform Gaussian smearing to each of the 3-momentum components to qualitatively simulate a 15\% mass resolution. For example, taking $m_S = 500 \ {\rm MeV}$ and implementing this invariant mass cut, the background is further reduced by a factor of ${\sim}8$ and we keep ${\sim}90\%$ of the signal (the remaining 10\% are events where the softer $\mu^-$ originates from $S$ decay). The requirements on achieving such an invariant mass resolution are discussed at length in Section~\ref{sec:experiment}.

\begin{figure}[t]
\centering
\includegraphics[width=0.46\textwidth]{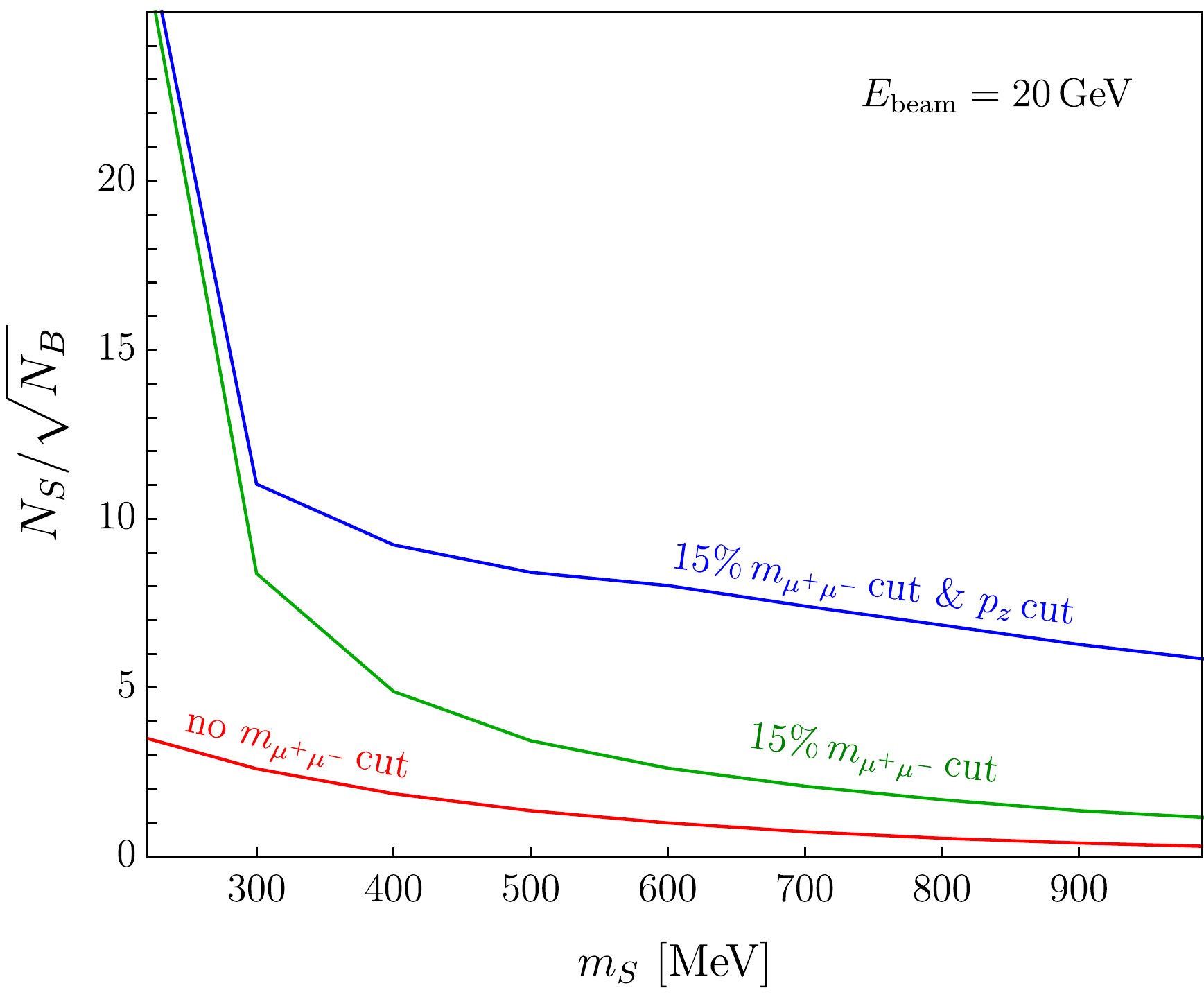}~~~~~~
\includegraphics[width=0.47\textwidth]{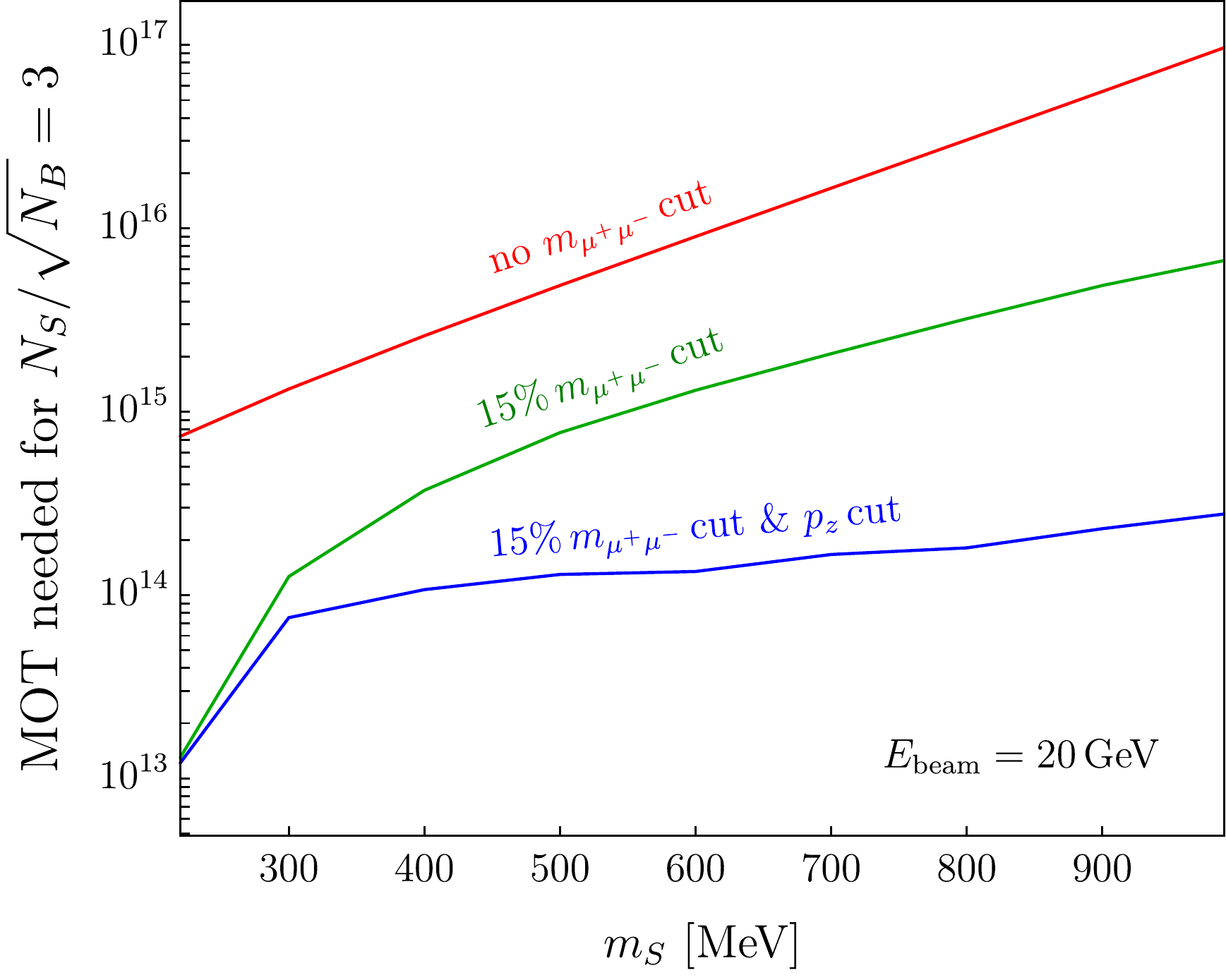}
\caption{\textbf{Left:} 
 Sensitivity as a function of scalar mass $m_S$ for various cuts, setting $g_S$ to the minimum value which resolves $\Delta a_\mu$ at each $m_S$ and assuming $10^{15}$ MOT and $\ell_T = 100 \ {\rm cm}$. \textbf{Right:} MOT required to reach $N_S/\sqrt{N_B} = 3$ as a function of $m_S$, for $g_S$ as in the left panel. Both panels assume a muon beam energy of 20 GeV.}
\label{fig:sensitivity}
\end{figure}

Figure~\ref{fig:sensitivity} (left) shows the effect on the sensitivity $N_S/\sqrt{N_B}$ from applying one or both cuts for a 20 GeV beam and $10^{15}$ MOT, using the same signal normalization that resolves $\Delta a_\mu$. We show the effect of an invariant mass cut alone (green curve) to emphasize the fact that a standard ``bump hunt'' strategy is much less efficient for our particular signal process without applying the $p_z$ cut. After applying both the invariant mass cut and the $p_z$ cut, the signal efficiency is $\sim 83 \%$ for low $S$ masses near 220 MeV and $\sim 50 \% $ for higher $S$ masses near 1 GeV. Due to the clear separation of signal and background kinematics, the combined effect of both cuts can improve the sensitivity by an order of magnitude for $m_S \lesssim 1 \ {\rm GeV}$, and even render the experiment free of irreducible QED background for larger masses. To illustrate the experimental requirements necessary to achieve full coverage of the $g-2$ parameter space for a given $m_S$, in Figure~\ref{fig:sensitivity} (right) we plot the muon flux required to achieve approximate $3\sigma$ discovery sensitivity, $N_S/\sqrt{N_B} = 3$, as a function of $m_S$. With $15\%$ invariant mass resolution and a 100 cm iron target region, we can probe the full parameter space that resolves the anomaly up to 1 GeV with $\simeq 3 \times 10^{14}$ MOT. For larger $m_S$ at this luminosity, the QED background becomes subdominant to the combinatorial background in a realistic experimental implementation, as we will describe in Sec.~\ref{sec:combinatorics} below.

\section{Potential reach at the SpinQuest Experiment}
\label{sec:experiment}

To make the discussion in Sec.~\ref{sec:calculations} concrete, and to illustrate the experimental considerations relevant for a practical implementation of our proposal, we imagine a setup similar to the SpinQuest experiment at Fermilab \cite{Apyan:2022tsd}, where a 120\,GeV proton beam impinges on a magnetized steel beam dump ($Z = 26$) many radiation lengths thick. Target interactions produce a spectrum of lower-energy secondary muons, which continue through the target and can produce $S$ through bremsstrahlung during any scattering process, as described above. To identify the vertex corresponding to $S$ decay, we consider only signal events which occur in the last portion of the 5\,m beam dump, of length $\ell_T \simeq 100 \ {\rm cm}$. As the simulated muon spectrum at proton beam dump experiments is still to be fully validated~\cite{Rella:2022len}, we leave a detailed sensitivity calculation for future work and consider a monochromatic 20 GeV muon beam. We note, however, that both signal and background cross sections are largely independent of the beam energy as shown in Fig.~\ref{fig:xsec}, so treating the muons as monochromatic is a decent approximation up to experimental acceptance effects.

\subsection{Muon beam properties}

\begin{figure}[t!]
    \centering
    \includegraphics[width=0.5\textwidth]{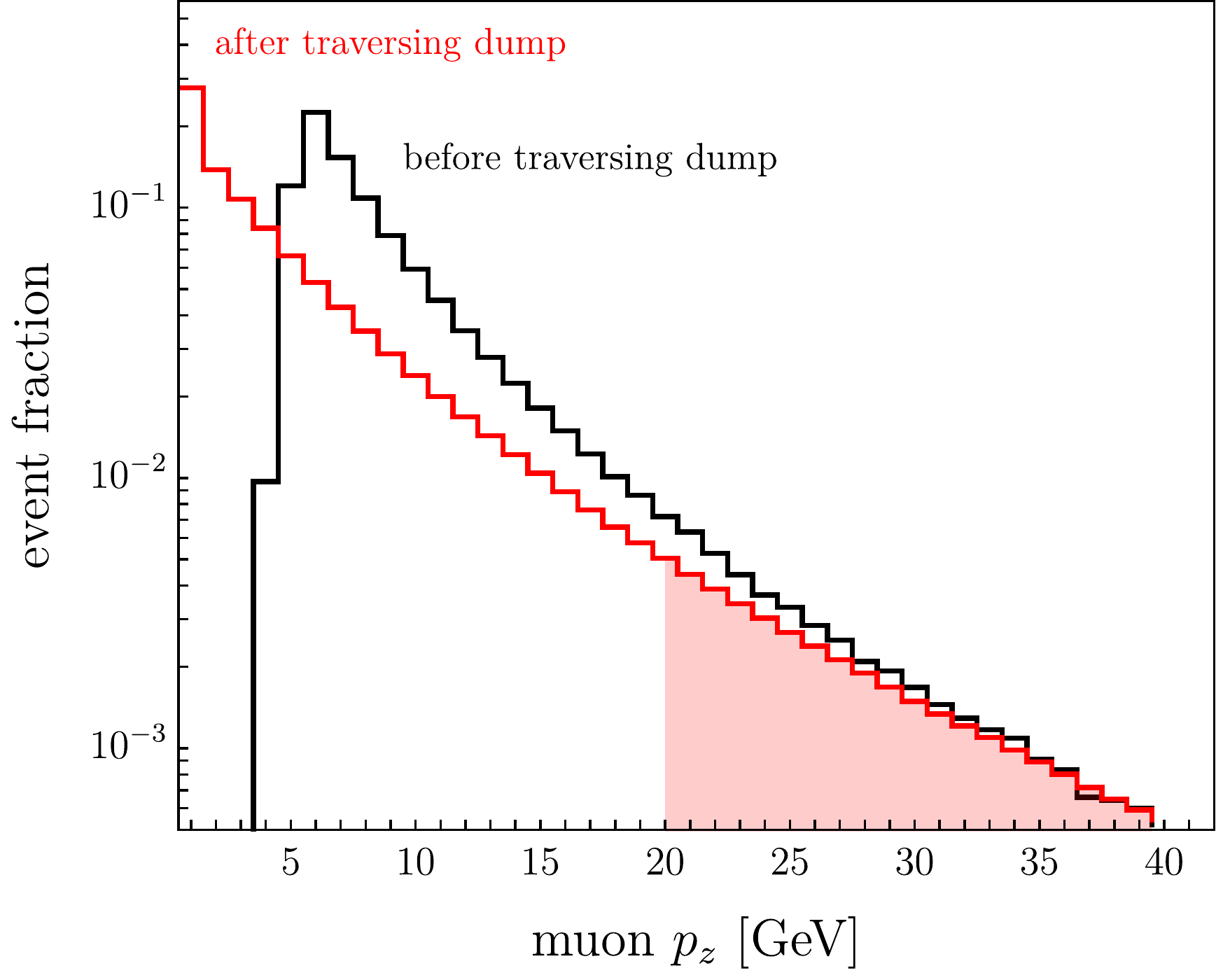}
\caption{Secondary muon momentum spectrum from a primary 120\,GeV proton beam. The part of the spectrum consisting of muons which exit the dump with $p_z > 20 \ {\rm GeV}$, which we treat as our muon beam, is shaded in red.
}
    \label{fig:pzComp}
\end{figure}

The signal sensitivity and estimated rate of backgrounds depend on the expected secondary muon momentum spectrum, following their traversal of the bulk of the beam dump. To facilitate these studies, we perform a toy simulation of muons passing through the dump, based on their initial spectra obtained from a full Geant4 simulation of the SpinQuest experimental setup~\cite{Apyan:2022tsd}.
Muons are randomly assigned to a position at the beginning of the dump based on the expected beam profile and propagated to the end of the dump, accounting for (a) initial muon momenta transverse to the beam, (b) small and wide-angle scattering within the dump, and (c) the effect of a 1.9\,T magnetic field.
The momenta of muons before and after the dump are shown in Figure~\ref{fig:pzComp}.
The relative composition of muons is approximately 55\% (45\%) for positively (negatively) charged muons.

To estimate the time required to achieve a given number of MOT at SpinQuest, we can use the finding of Ref.~\cite{Apyan:2022tsd} that 1.9 muons per RF bucket reach the first tracking station. Combining this with our simulation model, we find that 0.8 such muons have $p_z>20$~\,GeV. Each 4\,sec spill of 53\,MHz buckets thus yields $2\times10^8$~\,MoT with energy above 20 GeV, or $3\times10^{11}$~\,MoT in 24 hours of running.
Assuming that the accelerator complex can deliver beam with an average duty factor of 50\% over the full year, this amounts to $5 \times 10^{13}$~\,MOT accumulated per year of operation in the nominal configuration, such that our target luminosity for $m_S$ up to 1 GeV can be in principle achieved in 6 years of running.

\subsection{Reconstruction efficiencies and mass resolution}
\label{sec:massres}

The capability to efficiently trigger and accurately reconstruct di-muons will drive the sensitivity of muon spectrometer experiment to scalar decays.
The forward region of SpinQuest is instrumented with an emphasis on detecting muons from the decay of a high mass virtual photon or meson produced through target interactions.
Consequently, we consider only muons with momenta above 5\,GeV in the following analysis, assuming the trajectories of particles with lower momenta cannot be accurately measured.
While imperfect acceptance can lead to a small additional inefficiency, the effect has a complex dependence on the exact geometry of the detector and details of the reconstruction algorithms, which we do not attempt to reproduce in this work.
We note that despite the small loss in acceptance, a minimum requirement on the muon momentum removes background processes far more efficiently than signal, due to the soft sub-leading muon expected from radiative processes, as seen in Figures~\ref{fig:ch1} and ~\ref{fig:sensitivity}.

\begin{figure}[t]
\centering
\includegraphics[width=0.46\textwidth]{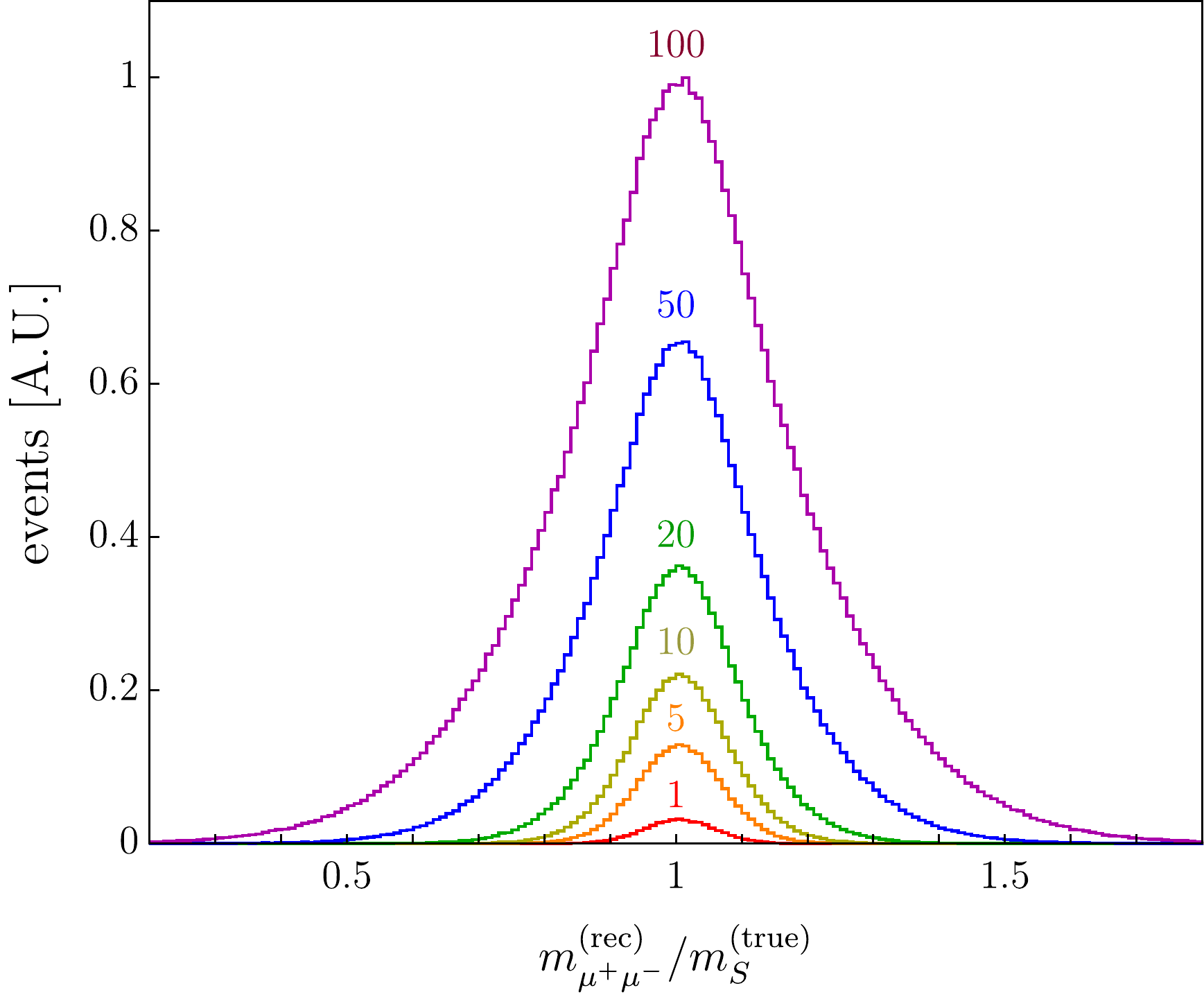}~~~
\includegraphics[width=0.455\textwidth]{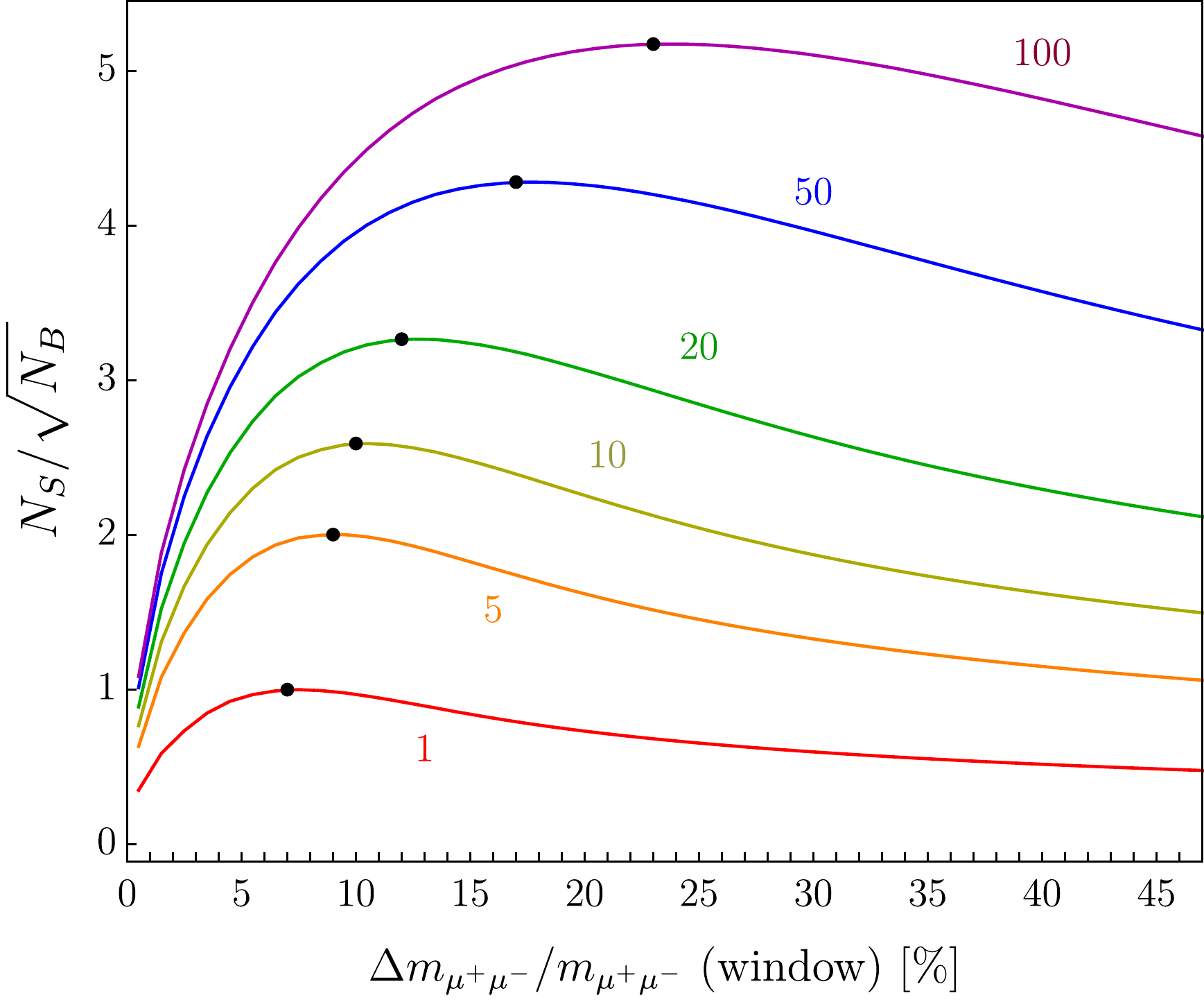}
\caption{\textbf{Left:} Reconstructed di-muon mass is shown for events with $m_S=1$\,GeV produced in the last portion of the dump, taking into account both multiple scattering effects and a 5\% experimental resolution.
Mass distributions are shown for several target thicknesses in units of $X_0$, with larger decay volumes increasing yields but degrading the mass resolution. Note that our fiducial value of $\ell_T = 100 \ {\rm cm}$ corresponds to about $57 X_0$ in steel.
\textbf{Right:} The corresponding signal sensitivities ($N_S/\sqrt{N_B}$) as a function of the mass window for each of the same thicknesses. The optimal window size is indicated with a black dot. A flat background in the neighborhood of the scalar mass is assumed.
\label{fig:targetDepth}
}
\end{figure}

Another key driver of the search sensitivity is the achievable di-muon mass resolution, which should be small to reduce the level of backgrounds, particularly for small scalar masses.
We assume a 5\% mass intrinsic experimental resolution, in accordance with the studies of Ref.~\cite{Apyan:2022tsd} based on decays between the dump and tracking stations.
However, because the scalar decays promptly after being produced within the last fraction of the dump, this mass resolution will generally be further degraded by multiple scattering.
Using the simulation described above, we propagated this effect to the mass resolution of $S\to\mu^+\mu^-$ pairs as a function of the target length traversed.
This leads to a trade-off, illustrated in Figure~\ref{fig:targetDepth}, where scalars produced earlier in the dump can be selected to enhance the overall signal rate, at the price of poorer mass resolution.
In this study, we take an effective target thickness of $\ell_T = 100 \ {\rm cm}$ as a representative choice, corresponding to an approximate 15\% total mass resolution including multiple-scattering effects. 
Finally, we need to be able to determine if the candidate dimuon mass pair originated from the target region. 
The displaced vertex resolution in the $z$ axis determined from simulation studies in Ref.~\cite{Apyan:2022tsd} is approximately 10\,cm.  This would be degraded by multiple scattering effects in the last ${\sim}1$\,m of the dump but would still be sufficient to determine if the dimuon pair is in the target region with reasonable efficiency.  

\subsection{Combinatorial background}
\label{sec:combinatorics}

In addition to the irreducible QED background discussed in Sec.~\ref{sec:calculations}, an additional experimental background can arise through combinations of independently-produced muons in the dump that conspire to produce an apparent di-muon vertex. This effect, unlike the irreducible radiative and trident backgrounds, is sensitive to details of the experimental setup such as vertex resolution, which we take as $\sigma_{x,y}=1$\,cm in line with SpinQuest's projected capabilities~\cite{Apyan:2022tsd}.

\begin{figure}[t!]
    \centering
    \includegraphics[width=0.49\textwidth]{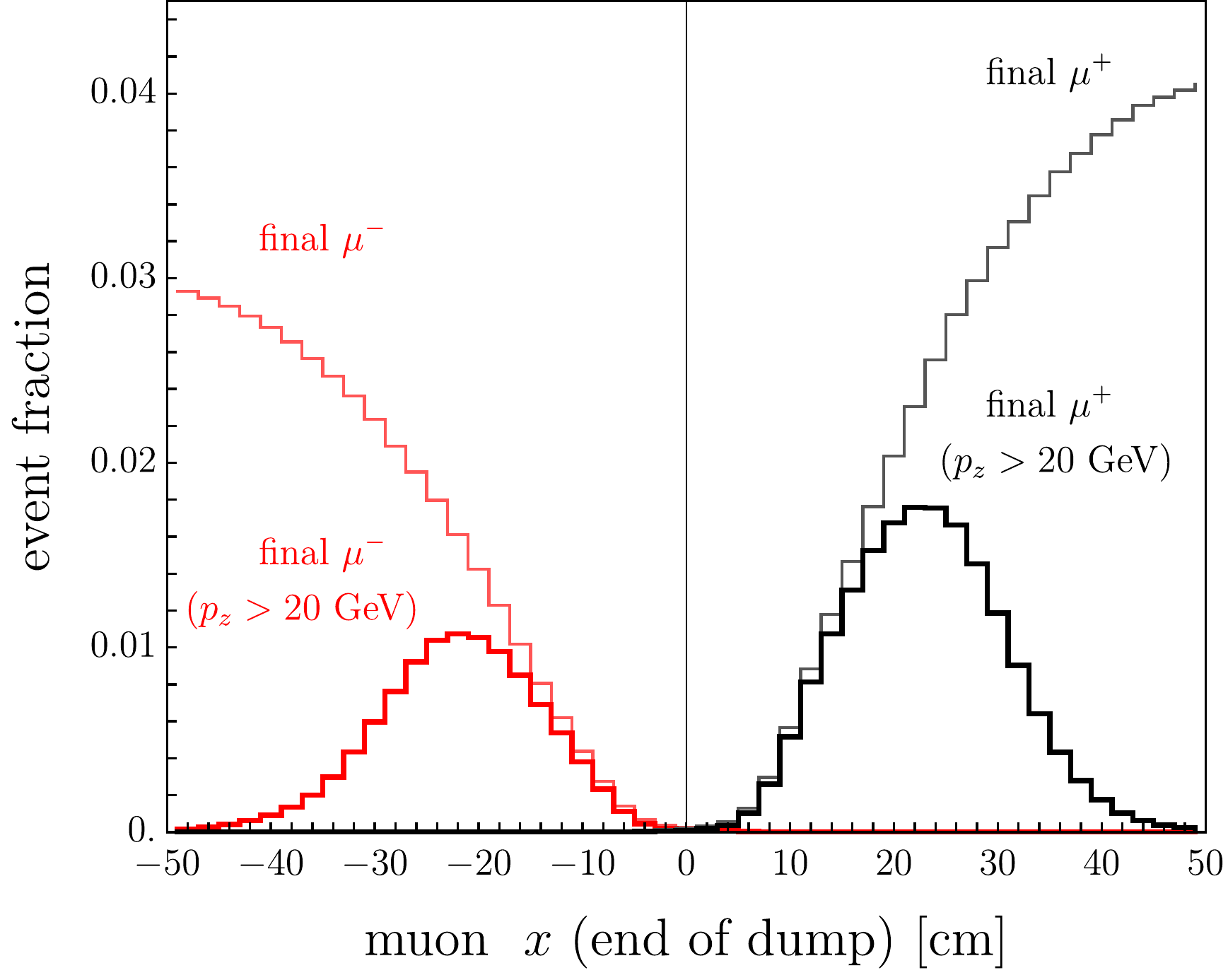}~~
    \includegraphics[width=0.49\textwidth]{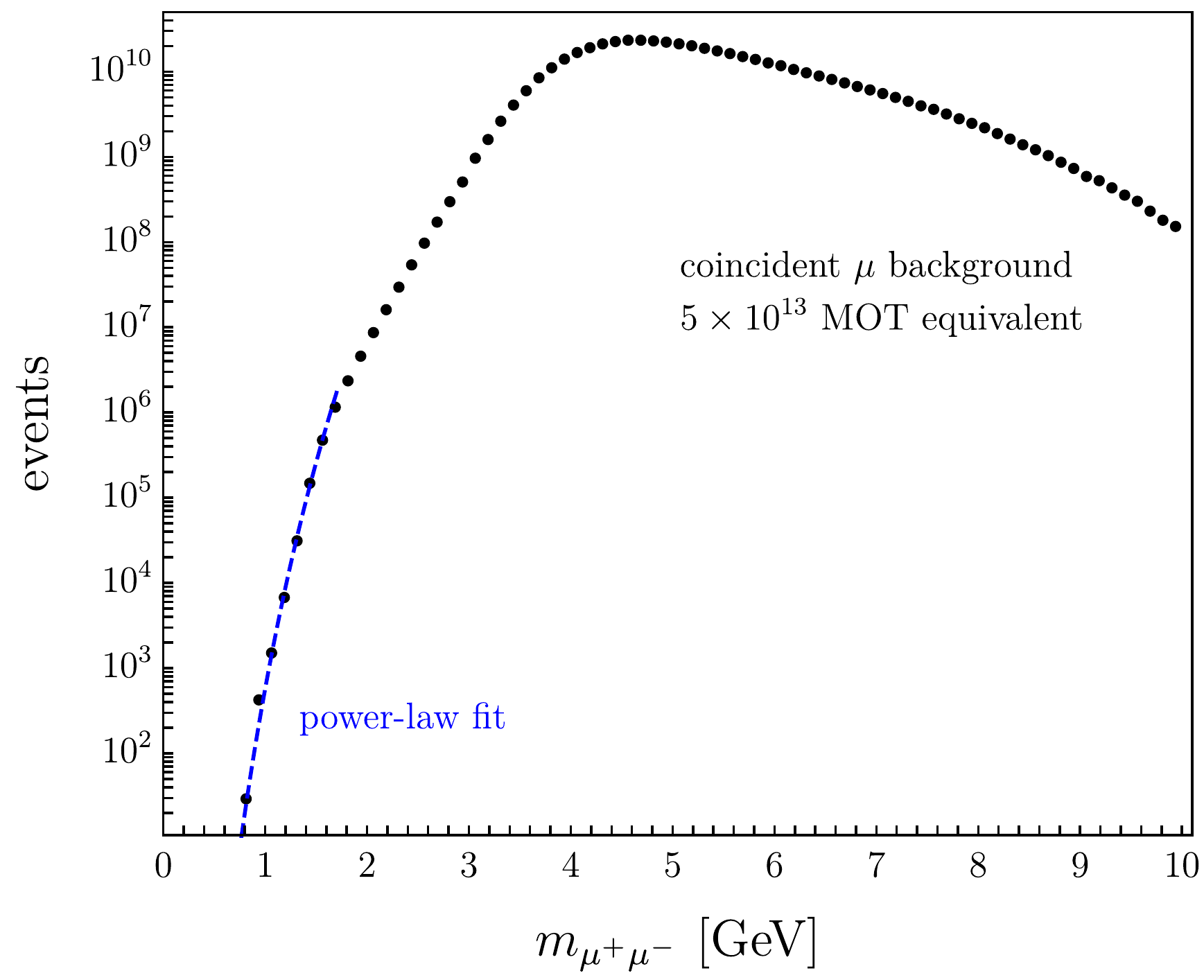}
\caption{\textbf{Left:} The position of positively and negatively charged muons in the bending plane, after traversing the beam dump. 
   \textbf{Right:} The $m_{\mu^-\mu^+}$ mass distribution for $\mu^-\mu^+$ pairs passing a common vertex requirement.}
    \label{fig:mumumass}
\end{figure}

The chance of finding a $\mu^-\mu^+$ vertex from uncorrelated muons is generally low because the magnetic field separates the oppositely-charged particles produced early in the dump.
Here the ``wrong-sign'' muon can only form a potential  $\mu^-\mu^+$ vertex if the combined effects of initial beam momenta and scattering in the dump are enough to compensate for the strong magnetic field bending its path in the opposing direction.
Figure~\ref{fig:mumumass} (left) shows the distribution of $\mu^+$ and $\mu^-$ positions in the bending plane at the end of the dump.
As $\mu^+S\to \mu^+\mu^-\mu^+$ signal events would result from the production of an $S$ off of a $\mu^+$ late in the dump, signal vertices will generally be concentrated at large $x$ (signal efficiency is $\epsilon_\text{sig}=95\%$ for $x_{\text{dump}}>10$\,cm), and vice-versa for $\mu^-S$ events.
Our simulation predicts a contamination rate of wrong-sign muons of $3\times10^{-4}$ after selecting events displaced by 10\,cm in the bending plane.

\begin{figure}[t]
\centering
\includegraphics[width=0.7\textwidth]{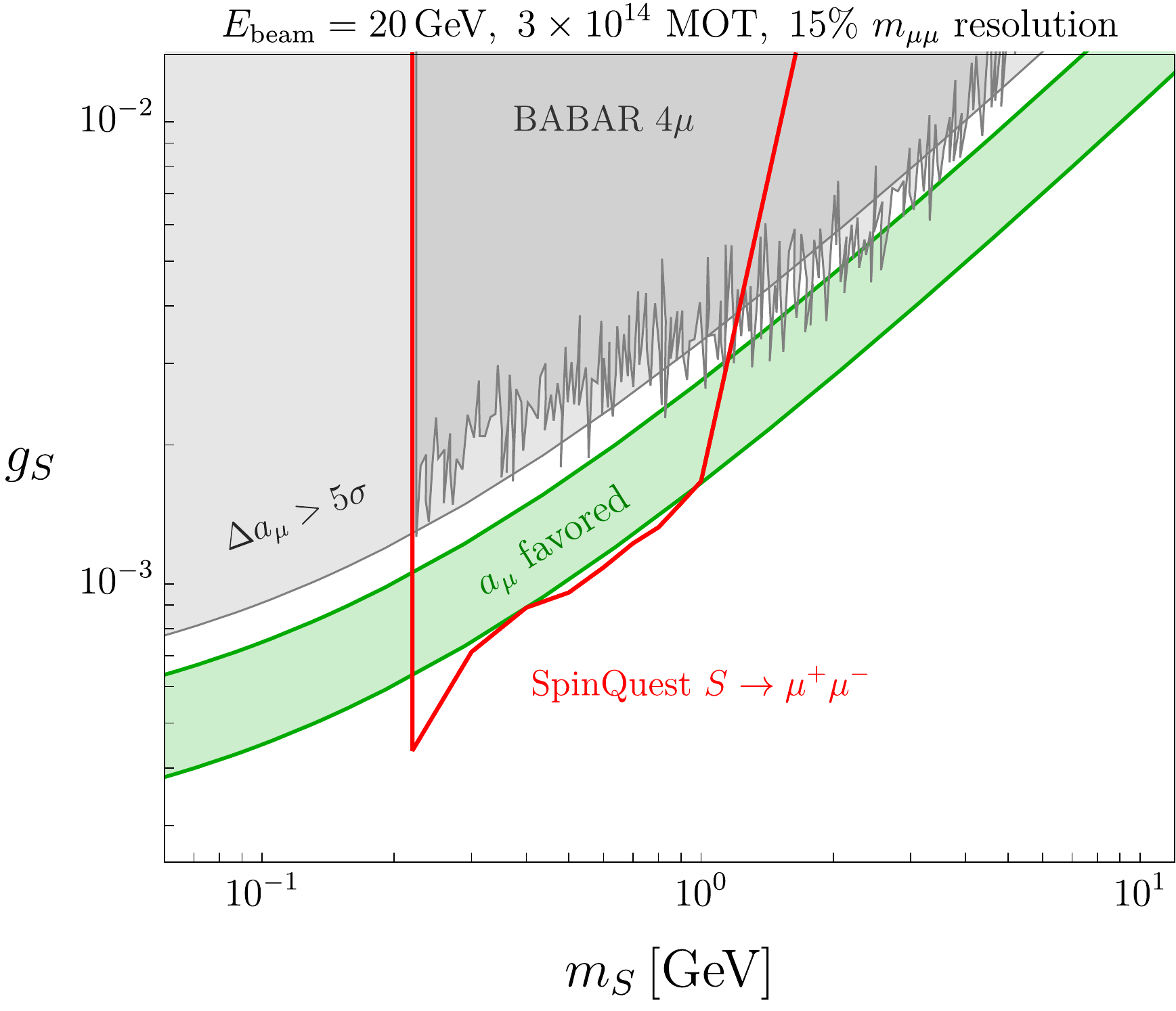}
\caption{ Sensitivity to the $(g-2)_\mu$ parameter space for a SpinQuest-like spectrometer at a proton beam dump experiments. We assume the same fiducial parameters as discussed in Sec.~\ref{sec:calculations}, namely 20 GeV beam energy, $\ell_T = 100 \ {\rm cm}$, 15\% invariant mass resolution, and $p_T > 5 \ {\rm GeV}$ on the hardest $\mu^+ \mu^-$ pair. We include both irreducible QED background and reducible combinatorial background. A muon flux of $3 \times 10^{14}$ MOT, corresponding to approximately 6 years of running at nominal SpinQuest luminosity, can fully cover the preferred region for $\Delta a_\mu$ to $2\sigma$ for $m_S < 1 \ {\rm GeV}$.
}
\label{fig:Reach}
\end{figure}

The rate of combinatorial $\mu^+\mu^-$ pairs can be computed from the rate of wrong-sign scatters into the signal-enriched region together with the expected position spread and vertex resolution.
The experimental resolution motivates a requirement that the reconstructed coordinates transverse to the beams agree to within 2\,cm, which occurs randomly in only 0.6\% of background events.
All together, this toy simulation predicts a rate of $5\times10^{-6}$ candidate pairs of 5\,GeV muons per each time sample.
The distribution of masses from these background events are shown in Figure~\ref{fig:mumumass} (right) for a sample of $5 \times10^{13}$ MoT.
Considering a 15\% window in di-muon mass, we expect this sample to contain about 100 combinatorial background events at 1 GeV. This background grows rapidly with mass (a power law fit yields a dependence $\sim m_{\mu^+ \mu^-}^{15}$), with $7 \times 10^{6}$ events for $m_S=2$\,GeV, illustrating the need for novel experimental strategies to probe the high-mass region.
Less than one combinatorial background event is predicted below 800\,MeV for this luminosity.

\subsection{Estimated sensitivity at SpinQuest}

Fig.~\ref{fig:Reach} shows our estimated sensitivity to the $g-2$ parameter space at a SpinQuest-like experiment, including the invariant mass resolution, target length, and combinatorial effects discussed above. Our limit on $g_S$ is defined%
\footnote{While systematic uncertainties are neglected in this analysis, they are expected to have negligible impact due to the robust background estimate from $m_{\mu^+\mu^-}$ side-band fits.  Background sculpting due to kinematic thresholds near $m_S \sim 2m_\mu$ may require careful treatment, which we will leave for future work.}
by solving $N_S/\sqrt{N_B} = 3$ for $g_S$ at each mass value $m_S$.
As anticipated, with $3 \times 10^{14}$ MOT, we can cover the full $\Delta a_\mu$ parameter space up to $2\sigma$ for $m_S < 1 \ {\rm GeV}$. We estimated the combinatorial background by scaling the results of Fig.~\ref{fig:mumumass}, but the extremely steep scaling with mass results in a sharp degradation of the sensitivity of our proposed experiment at around 1 GeV. We have verified that the sensitivity is largely unchanged for different beam energies; the same muon flux with 30 GeV or 40 GeV muons can probe the same parameter space, which suggests that our estimates are likely robust to the precise secondary muon spectrum at proton beam dump experiments. 

\section{Conclusions and Outlook}
\label{sec:outlook}
In this paper, we have introduced a new search strategy for muonphilic particles using proton beam dump spectrometers. 
As primary protons impinge on a fixed target, they produce a beam of secondary muons whose interactions with the target material
can produce new muonphilic particles. If these new states decay visibly to dimuons, the daughter particles
emerge from the target region and their combined invariant mass can be reconstructed using a downstream tracking station. This search
strategy can be parasitically executed at the Fermilab SpinQuest experiment and
can achieve unprecedented sensitivity to 
new muonphilic particles with $3 \times 10^{14}$ muons on target and appropriate
analysis cuts.

While the search strategy we outlines here is general for any muonphilic particles, our discussion has been framed around particles that resolve the longstanding 
muon $g-2$ anomaly, which is arguably the longest-standing disagreement between SM predictions and experimental measurements. Assuming the theoretical prediction of $g-2$ remains unchanged with the inclusion of recent lattice QCD results, all possible beyond-the-SM solutions should be tested comprehensively. Here we have found that a proton beam dump spectrometer can cover parameter space in a highly complementary region to missing-momentum experiments such as $M^3$~\cite{Kahn:2018cqs}, which can probe $m_S < 2m_\mu$, and NA64-$\mu$~\cite{Sieber:2021fue}, which can fully probe the parameter space for an $L_\mu - L_\tau$ gauge boson. Future $B$-factories such as Belle-II can also cover the visible scalar decay parameter space~\cite{Capdevilla:2021kcf}, but the full luminosity may be a decade away. In the intervening years, our analysis shows that proton beam dump experiments such as SpinQuest can potentially discover the new physics responsible for $\Delta a_\mu$ at masses below 1 GeV in a reasonable ${\sim}6$ years of running, and likewise has sensitivity to other muonphilic particles in this mass range.

More detailed analyses by experimental collaborations are required to produced a more refined sensitivity projection, including in-situ measurements of the SpinQuest muon spectrum and several detector effects such as detector reconstruction efficiency and trigger efficiency.  Additional and more sophisticated analysis techniques include multivariate kinematic selections and an optimization of the beam dump target region.  
Nonetheless, even with such considerations, the sensitivity of the SpinQuest experiment can benefit even more significantly from beamline considerations and detector improvements.  Increasing the duty factor of the SpinQuest experiment, which currently takes data for 4\,s out of 1\,minute, could increase the expected MOT in a year by approximately an order of magnitude -- of course at the expense of other experiments.  Furthermore, the RF bucket-to-bucket intensity is limited by the efficiency of the slow extraction of the beam from the main injector and the occupancy of the detector.  By improving the proton beam intensity and using more modern highly-granular tracking detectors, especially at the front of the spectrometer, the sensitivity of the experiment could also be drastically improved.

\vspace{0.5cm}
\noindent \textbf{Acknowledgments}: We thank Christina Gao, Yongbin Feng, Philip Harris, and Yiming Zhong for helpful conversations. DF and YK especially thank Dmitriy Kirpichnikov for invaluable assistance with CalcHEP. We thank the dark sectors community from the SpinQuest collaboration for their feedback. The work of DF and YK was supported in part by DOE grant DE-SC0015655.  The work of CH, GK, CMS, and NT is supported by Fermi Research Alliance, LLC under Contract No. DE-AC02-07CH11359 with the U.S. Department of Energy, Office of Science, Office of High Energy Physics and the Fermilab Lab Directed R\&D (LDRD) program.

\bibliography{bibliography}

\end{document}